 \documentstyle[emulateapj]{article}

\submitted{accepted ApJS, Aug 2000, in press}

\begin{document}
\title{Magellanic-Cloud-Type Interstellar Dust Along Low Density Sightlines 
in the Galaxy} 

\author{Geoffrey C. Clayton$^1$, Karl D. Gordon$^2$, and Michael J. Wolff$^3$}

\altaffiltext{1}{Department of Physics \& Astronomy, Louisiana State 
University \\
   Baton Rouge, LA 70803 \\ Email: gclayton@fenway.phys.lsu.edu}
\altaffiltext{2}{Steward Observatory, University of Arizona, Tucson, AZ 85721 \\ 
Email: kgordon@as.arizona.edu}
\altaffiltext{3}{Space Science Institute, 1540 30th Street, Suite 23
Boulder, CO 80303-1012  \\ Email: wolff@colorado.edu}

\begin{abstract} 
We have studied the UV extinction properties along 30 Galactic
sightlines using data from the {\it International Ultraviolet
Explorer} (IUE) archive that have never been previously examined.
These distant (d $>$ 1 kpc) sightlines were selected to investigate
the distribution and physical conditions of gas located in low density
regions of the Galactic disk and halo.  The average densities along
these sightlines are extremely low.  It is likely that they are
dominated by the warm intercloud medium and have little contribution
from the cold cloud medium.  We find that a subsample of these
sightlines has extinction curves with weak bumps and very steep far-UV
extinction reminiscent of the Magellanic clouds.  These sightlines all
lie in the region bounded by 325$^o \leq l \leq 0^o$ and -5$^o \geq b
\geq -11^o$.  The gas along these sightlines shows forbidden
velocities which may indicate that the dust has been subject to
shocks.  This type of low density sightline may mimic the environments
found in the Magellanic Clouds.  Large values of N(Ca II)/N(Na I)
indicating low depletion are associated with steep far-UV extinction.
A possible correlation exists between decreasing bump strength and
increasing far-UV steepness for extinction curves in the Galaxy and
the Magellanic Clouds.
\end{abstract}

\section{Introduction} 

There is an average Milky Way extinction relation, A($\lambda$)/A(V),
over the wavelength range 0.125 $\mu$m to 3.5 $\mu$m, which is
applicable to a wide range of interstellar dust environments,
including lines of sight through diffuse dust, and dark cloud dust, as
well as dust associated with star formation (Cardelli, Clayton, \&
Mathis 1989 (CCM); Cardelli \& Clayton 1991; Mathis \& Cardelli 1992;
Fitzpatrick 1999).  The existence of this relation, valid over a large
wavelength interval, suggests that the environmental processes which
modify the grains are efficient and affect all grains.  The CCM
relation depends on only one parameter, the ratio of
total-to-selective extinction, R$_V$ which is a crude measure of the
size distribution of interstellar dust grains.

However, the CCM relation does not appear to apply beyond the Milky
Way.  It does not always fit the observed extinction along sightlines
observed in the Magellanic Clouds and M31 (e.g., Clayton \& Martin
1985; Fitzpatrick 1985, 1986; Clayton et al.  1996; Bianchi et al.
1996; Gordon \& Clayton 1998; Misselt, Clayton, \& Gordon 1999).  The
2175 \AA~bump is weaker and the far-UV extinction is steeper in many
of the Magellanic cloud sightlines but there are also sightlines in
both the LMC and SMC where the dust extinction does follow CCM.  The
few lines of sight studied in M31 seem to show a CCM-like far-UV
extinction and a weak 2175 \AA~bump (Bianchi et al.  1996).  On the
other hand, the starburst nucleus of M33 appears to be associated with
Milky-Way-type dust (Gordon et al.  1999).  The variations in
extinction properties seen in the Magellanic Clouds and M31 may be due
to several factors.  Different environments, such as star formation
regions where large amounts of UV radiation and shocks are present,
may play a large role in processing dust.  Evidence for this can be
seen in the LMC where two distinct wavelength dependences of UV
extinction have been found for dust inside and outside the supergiant
shell, LMC 2, which lies on the southeast side of 30 Dor.  This
structure was formed by the combined stellar winds and supernovae
explosions from the stellar association at its center (Misselt et al.
1999).  In the SMC, the dust properties are even more extreme, showing
extinction curves for three of four sightlines which have virtually no
bump and are very steep in the far-UV (Gordon \& Clayton 1998).
Although the dust responsible for these curves is located near regions
of star formation in the SMC, the environment is likely to be less
severe than for the LMC 2 dust.  The 30 Dor region, where LMC 2 is
located, is a much larger star forming region than any in the SMC.
The dust environments in starburst galaxies and QSOs, which also show
SMC-like extinction, are much more extreme than 30 Dor (e.g., Gordon,
Calzetti, \& Witt 1997; Pitman, Clayton \& Gordon 2000; Gordon, Smith
\& Clayton 2000).  The SMC has star formation occurring at only 1\%
the rate of a starburst galaxy so other factors such as the known
differences in metallicity between galaxies may be important
(Fitzpatrick 1986; Gordon \& Clayton 1998; Misselt et al. 1999).

Setting aside global metallicity differences, are there sightlines in the Galaxy
where the dust environment is similar to those seen in the Magellanic clouds?  Real
deviations from CCM are seen in the Galaxy but deviations of the kind seen in the
Magellanic clouds have been seen only rarely (Cardelli \& Clayton 1991; Mathis \&
Cardelli 1992).  A few sightlines (e.g., 62542, 204827, and 210121) show
weak bumps and anomalously strong far-UV extinction for their measured values of
$R_V$.  Their extinction curves are plotted in Figure 1.  These deviant sightlines
represent a variety of dust environments.  The Galactic sightline toward HD
62542 is somewhat similar to LMC 2.  Its dust was swept up by bubbles blown by two
nearby O stars (Cardelli \& Savage 1988).  HD 204827 is also in a star formation
region where the dust has been subject to shocks (Clayton \& Fitzpatrick 1987).
HD 210121 lies
behind a single cloud in the halo.  There is no present activity near this cloud
although it was ejected into the halo at some time in the past.  There are
some important differences between these Galactic extinction curves
and those in the
Magellanic clouds.  The bump seen for HD 62542 is not just weak but
it is very broad and shifted to the blue (Cardelli \& Savage 1988).  
Mantles on the bump grains has been suggested as the reason for the weak,
broad, and shifted Galactic bumps (Mathis \& Cardelli 1992; Mathis 1994).  These
sightlines show that dust in a variety of environments with a range of
$R_V$ values can have extinction curves similar to those in the LMC.  However, none
of the anomalous Galactic sightlines, seen in Figure 1, approach the SMC extinction
properties.  The SMC dust has weaker bumps and steeper far-UV extinction than any
known Galactic or LMC sightline.

\vspace*{0.05in}
\begin{center}
\plotone{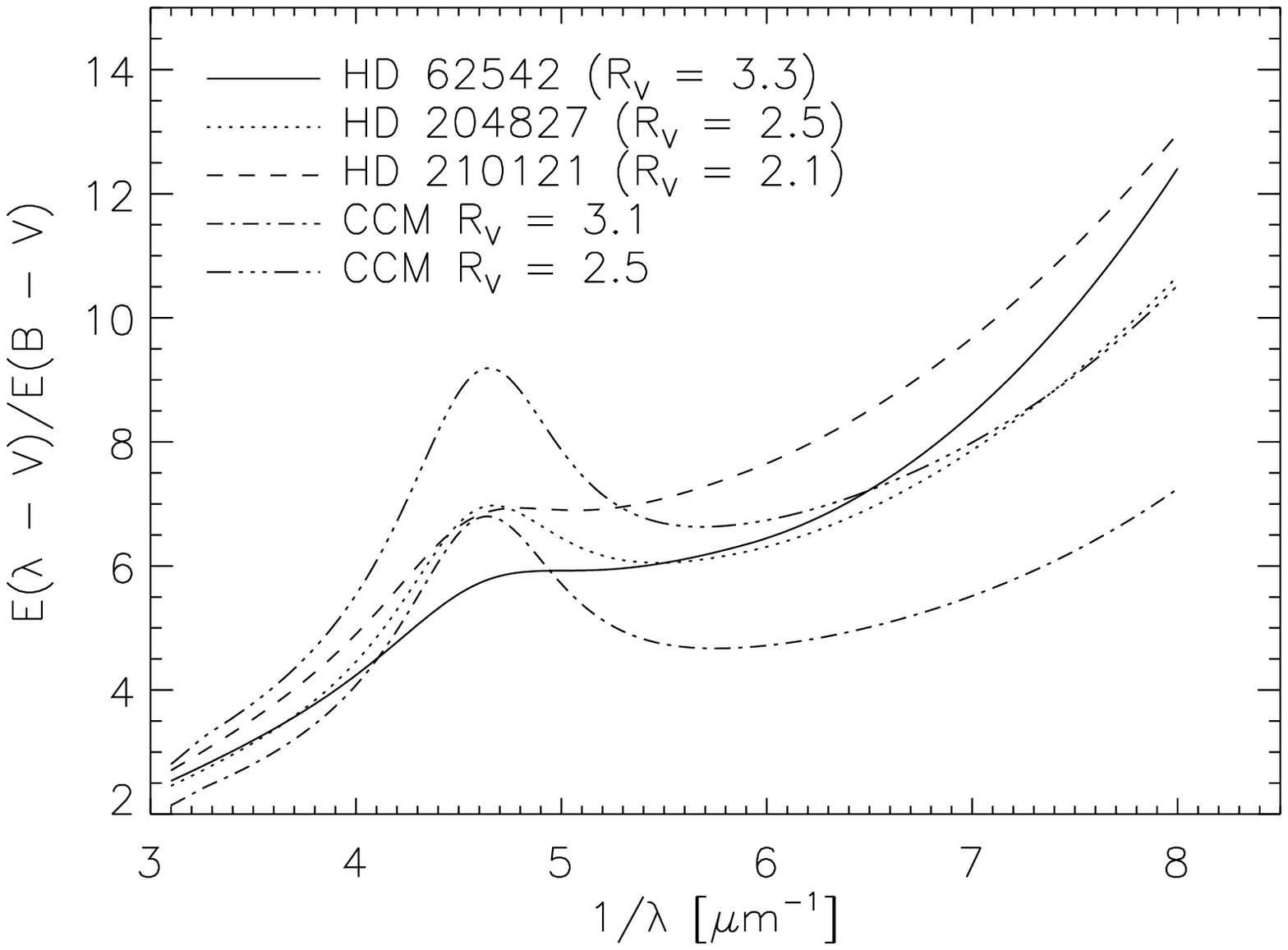}
\end{center}
\figurenum{1}
\figcaption{Sightlines in the Galaxy showing anomalous extinction. CCM curves for 
$R_V$ values of 2.5 and 3.1 are plotted for comparison.}
\vspace*{0.2in}

Most of the Galactic sightlines, that have been studied previously, 
differ in one respect from the LMC and SMC
sightlines.  They are significantly more reddened than the Magellanic cloud
sightlines.  In particular, those sightlines showing the greatest deviations from
CCM, those near the supershell LMC 2 and those in the SMC, all have E(B-V) $<$
0.25.  Of the twenty-nine CCM sightlines only two have E(B-V) $<$ 0.30.  The others
range up to E(B-V) = 1.2.  Similarly, the Fitzpatrick \& Massa sample of eighty
stars includes only seven with E(B-V) $<$ 0.30 (Fitzpatrick \& Massa 1990;
Fitzpatrick 1999).  Therefore, the dust along the Magellanic cloud sightlines is
more diffuse and more representative of the warm intercloud medium than the cold
cloud medium which is better represented in the Galactic samples.

Kiszkurno-Koziej \& Lequeux (1987) suggest from ANS extinction measurements of 1200
stars in the Galaxy that there may be a correlation between UV extinction
parameters and distance from the Galactic plane.  As $\vert z \vert$ increases, the bump becomes
weaker and the far-UV extinction stronger.  These sightlines have low reddenings and
long sightlines so they are also more diffuse and therefore more like those in 
the Magellanic
clouds.  To investigate whether the extinction properties observed in the
Magellanic clouds are related to the diffuse nature of the sightlines, a sample of
long sightlines with low reddenings 
in the Galaxy was chosen and UV data were obtained with the 
{\it International Ultraviolet Explorer} (IUE) so that
extinction curves could be constructed.

\section{The Sample}

Sembach, Danks, \& Savage (1993) obtained high resolution Na I D and Ca II K
spectra for a sample of distant (d $>$ 1 kpc) stars.  These sightlines were
selected to investigate the distribution and physical conditions of gas located in
low density regions of the Galactic disk and halo.  The sightlines listed in Table
1 were selected from the Sembach et al.  sample for a complementary study of the UV
extinction properties of interstellar dust in low density conditions.  The Sembach
et al.  sample is limited to stars with spectral types between O8 and B3 which
makes them ideal for extinction studies.  Following the definitions of Sembach et
al.  the sightlines in Table 1 lie outside the Sagittarius spiral arm (IA1),
between the Sagittarius and Scutum-Crux spiral arms (IA2), beyond the Scutum-Crux
arm toward the Galactic center (GC), and the inner 4 kpc of the Galaxy
(IGC).  The stars lie at distances ranging from 1.5 to 9.5 kpc and have heights
above or below the plane of 0 to 1.5 kpc.  Twelve of the sightlines extend
into the Galactic halo, defined as $\vert z \vert$ $>$ 500 pc .  The locations of
these stars in the Galaxy are plotted in Figure 1 of Sembach et al.  (1993).

\begin{deluxetable}{llllllllllc}
\tablewidth{0pt}
\footnotesize
\tablecaption{Low Density Interstellar Sightlines in the Galaxy
\label{obsdat}}
\tablehead{
\multicolumn{1}{c}{$HD$} &
\multicolumn{1}{c}{$SpT$} &
\multicolumn{1}{c}{$V$} &
\multicolumn{1}{c}{$E_{B-V}$} &
\multicolumn{1}{c}{$l$} &
\multicolumn{1}{c}{$b$}&
\multicolumn{1}{c}{$d$}&
\multicolumn{1}{c}{$z$}&
\multicolumn{1}{c}{$Type$}&
\multicolumn{1}{c}{$n_o(H I)$}&
\multicolumn{1}{c}{$N(Ca II)/N(Na I)$}
}
\startdata
HD 64219&B2.5 III&9.72&0.17&241.99&0.94&3.46&0.06&IA1&0.098&0.49\nl
HD 69106&B0.5 IVnn&7.13&0.18&254.52&-1.33&1.49&-0.03&IA1&0.235&0.21\nl
HD 93827&B1 Ibn&9.31&0.23&288.55&-1.54&8.31&-0.22&IA2&0.064&0.45\nl
HD 94493&B1 Ib&7.23&0.20&289.01&-1.18&3.33&-0.07&IA1&0.121&0.66\nl
HD 97848&O8 V&8.68&0.30&290.74&1.53&2.69&0.09&IA2&0.225&0.50\nl
HD 100276&B0.5 Ib&7.16&0.26&293.31&0.77&2.96&0.04&IA2&0.172&0.38\nl
HD 103779&B0.5 Iab&7.20&0.21&296.85&-1.02&4.02&-0.07&IA2&0.104&0.54\nl
HD 104683&B1 Ib&7.92&0.19&297.74&-1.97&4.64&-0.16&IA2&0.090&0.79\nl
HD 104705&B0 Ib&7.76&0.26&297.45&-0.34&3.90&-0.02&IA2&0.128&0.67\nl
HD 113012&B0.2 Ib&8.12&0.34&304.21&2.77&4.11&0.20&IA2&0.188&0.50\nl
HD 148422&B1 Ia&8.60&0.28&329.92&-5.60&8.84&-0.86&GC&0.125&0.72\nl
HD 151805&B1 Ib&8.91&0.32&343.20&1.59&5.94&0.16&IGC&0.119&0.34\nl
HD 151990&O9 IV&9.46&0.39&334.99&-5.54&4.47&-0.43&GC&0.244&0.69\nl
HD 158243&B1 Iab:&8.15&0.19&337.59&-10.64&6.49&-1.20&IGC&0.145&0.63\nl
HD 160993&B1 Iab:&7.73&0.21&345.61&-8.56&5.20&-0.77&IGC&0.149&0.63\nl
HD 161653&B1 II&7.11&0.26&352.42&-5.26&1.82&-0.17&IA2&0.315&\nodata\nl
HD 163522&B1 Ia&8.46&0.19&349.57&-9.09&9.42&-1.49&IGC&0.119&1.11\nl
HD 164019&O9.5 III&9.26&0.53&1.91&-2.62&3.83&-0.28&GC&0.308&\nodata\nl
HD 164340&B0.2 III&9.25&0.15&352.06&-8.60&5.46&-0.82&IGC&0.105&0.95\nl
HD 165582&B1 II&9.33&0.24&357.49&-6.96&5.21&-0.63&IGC&0.152&0.85\nl
HD 167402&B0 Ib&8.95&0.23&2.26&-6.39&7.04&-0.78&IGC&0.121&0.79\nl
HD 168941&O9.5 II-III&9.34&0.37&5.82&-6.31&5.79&-0.64&IGC&0.212&$<$0.38\nl
HD 172140&B0.5 III&9.96&0.22&5.28&-10.61&6.85&-1.26&IGC&0.165&0.82\nl
HD 177989&B0 III&9.33&0.25&17.81&-11.88&4.91&-1.01&GC&0.223&$<$0.25\nl
HD 178487&B0 Ia&8.66&0.40&25.78&-8.56&7.66&-1.14&IGC&0.249&$<$0.31\nl
HD 179407&B0.5 Ib&9.41&0.31&24.02&-10.40&7.76&-1.40&IGC&0.224&$<$0.39\nl
\enddata
\end {deluxetable}

\section{Extinction Curves} 

Low dispersion short and long wavelength IUE spectra
were obtained between 1991 and 1994 by Jason Cardelli.  The spectra listed in Table
2 were downloaded from the IUE archive.  The archive spectra were reduced using
NEWSIPS and then were recalibrated using the method developed by Massa \&
Fitzpatrick (2000).  The short and long wavelength spectra for each star were
co-added, binned to the instrumental resolution ($\sim$ 5 \AA) and merged at the
maximum wavelength of the short wavelength spectrum.

\begin{deluxetable}{lllcll}
\tablewidth{0pt}
\small
\tablecaption{Observations}
\tablehead{\colhead{Star} & \colhead{Standard} & \colhead{SpT} & 
           \colhead{$\Delta$(B-V)} & \colhead{SWP spectra} &
           \colhead{LWR/LWP spectra}}
\startdata
HD 64219 & HD 31726 & B1V & 0.20 & SWP 47546 & LWP 25410/25412 \nl
HD 69106 & HD 63922 & B0III & 0.21 & SWP 11114/47547 & LWR 8825 \nl
HD 93827 & HD 62747 & B1.5III & 0.29 & SWP 48392/48400 & LWP 26160 \nl
HD 94493 & HD 119159 & B0.5III & 0.29 & SWP 47548/47571 & 
   LWP 25413/25420 \nl
HD 97848 & HD 47839 & O7V & 0.29 & SWP 39231 & LWP 18380 \nl
HD 100276 & HD 64760 & B0.5Ib & 0.25 & SWP 39256 & LWP 18397 \nl
HD 103779 & HD 64760 & B0.5Ib & 0.22 & SWP 48391/48399 & LWP 26159/26170 \nl
HD 104683 & HD 119159 & B0.5III & 0.28 & SWP 47560 & 
   LWP 25414/25424 \nl
HD 104705 & HD 63922 & B0III & 0.28 & SWP 39257 & LWP 18398 \nl
HD 113012 & HD 63922 & B0III & 0.41 & SWP 47559 & LWP 25423 \nl
HD 148422 & HD 64760 & B0.5Ib & 0.30 & SWP 48329 & LWP 26100 \nl
HD 151805 & HD 40111 & B1Ib & 0.34 & SWP 39267 & LWP 18408 \nl
HD 151990 & HD 188209 & O9.5Ia & 0.34 & SWP 47555/47570 & LWP 25419 \nl
HD 158243 & HD 91316 & B1Iab & 0.19 & SWP 48328 & LWP 26099 \nl
HD 160993 & HD 40111 & B1Ib & 0.21 & SWP 48334 & 
   LWP 26105/26109 \nl
HD 161653 & HD 62747 & B1.5III & 0.25 & SWP 47542 & LWP 25406 \nl
HD 163522 & HD 91316 & B1Iab & 0.19 & SWP 48335 & LWP 26106 \nl
HD 164019 & HD 167756 & B0Ia & 0.45 & SWP 33407/47543 & 
   LWP 13140/13141/25416/25417 \nl
HD 164340 & HD 119159 & B0.5III & 0.14 & SWP 47541/47552 & LWP 25441 \nl
HD 165582 & HD 119159 & B0.5III & 0.28 & SWP 48336 & LWP 26107 \nl
HD 167402 & HD 167756 & B0Ia & 0.20 & SWP 42336 & LWP 21095 \nl
HD 168941 & HD 63922 & B0III & 0.34 & SWP 33409/47540 & 
   LWP 13142/25404 \nl
HD 172140 & HD 119159 & B0.5III & 0.24 & SWP 48341 & LWP 26110/26172 \nl
HD 177989 & HD 119159 & B0.5III & 0.23 & SWP 42342 & LWP 21110 \nl
HD 178487 & HD 167756 & B0Ia & 0.38 & SWP 48413/48415 & LWP 26186 \nl
HD 179407 & HD 150168 & B1Ia & 0.28 & SWP 42334 & LWP 21093/21096 \nl
\enddata
\end{deluxetable}

\begin{figure*}[tbp]
\figurenum{2}
\plottwo{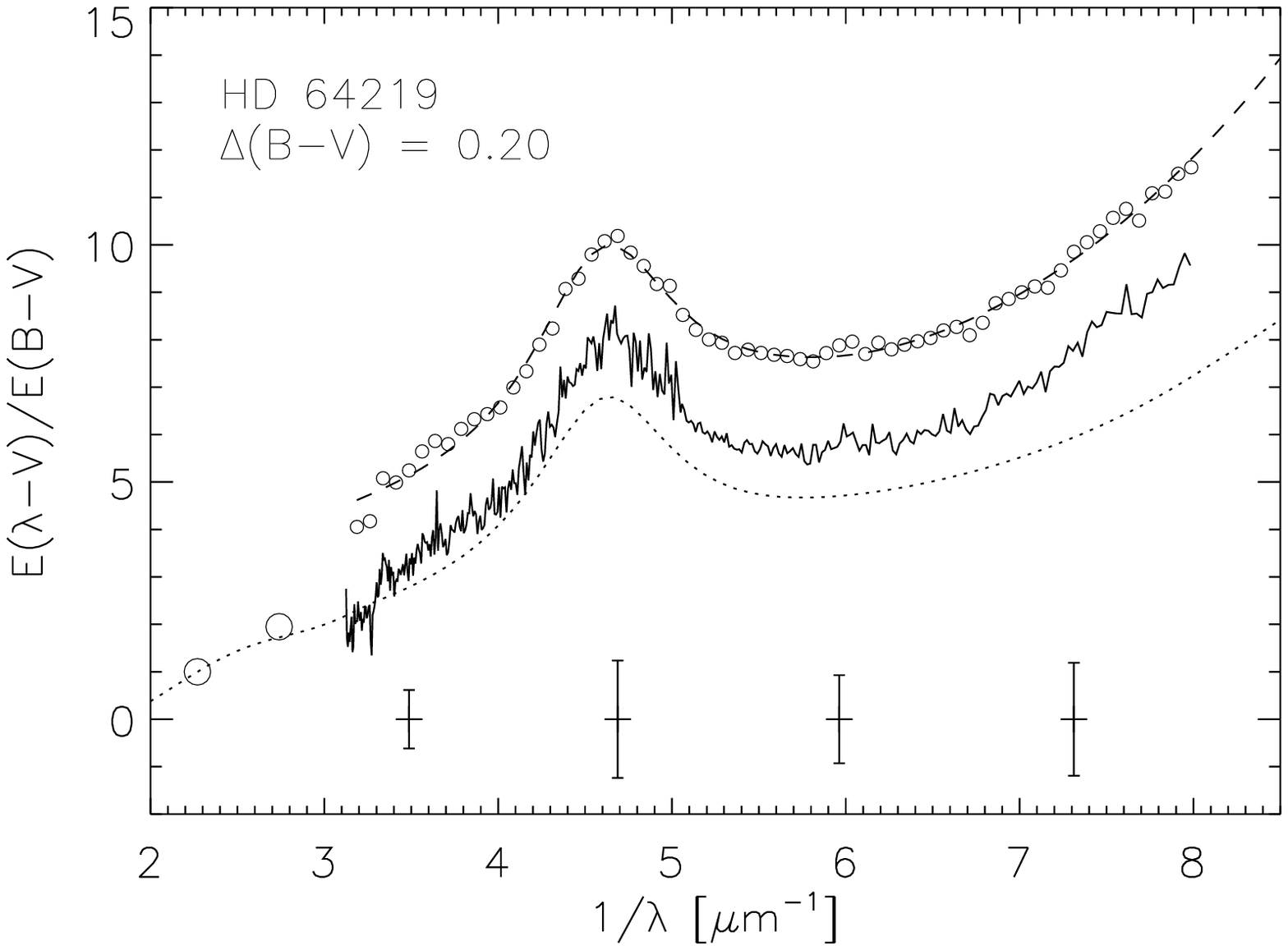}{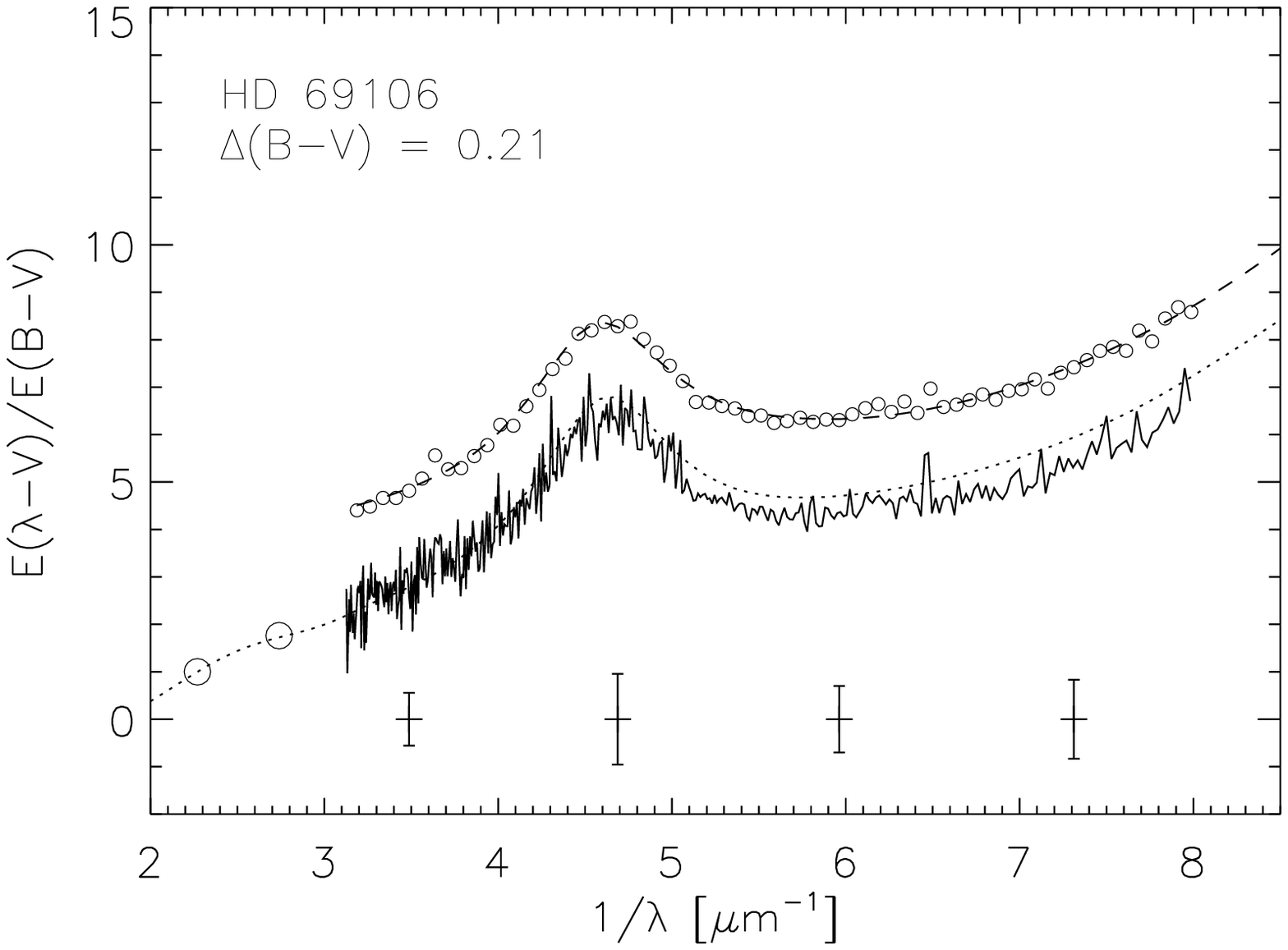} \\
\plottwo{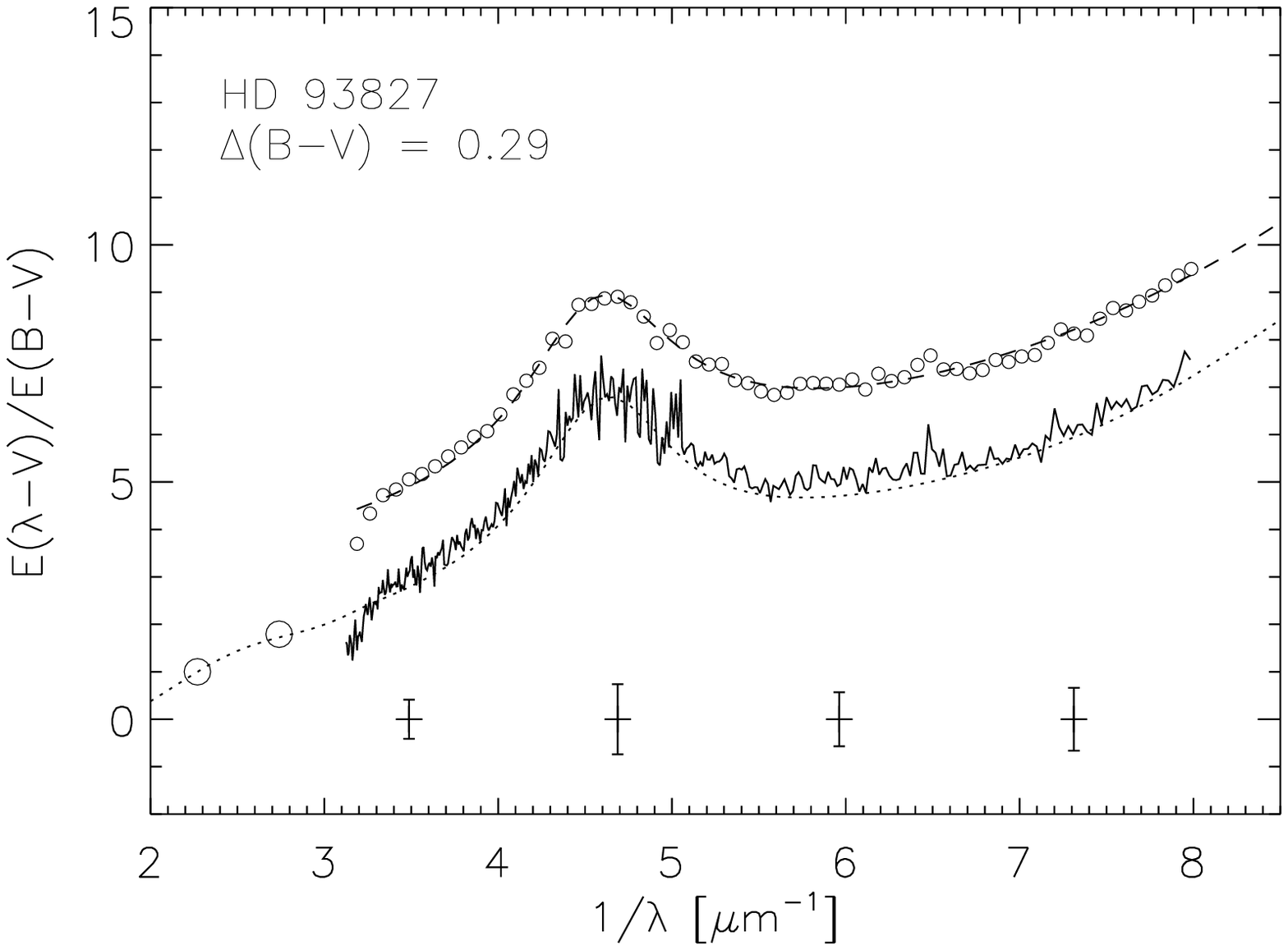}{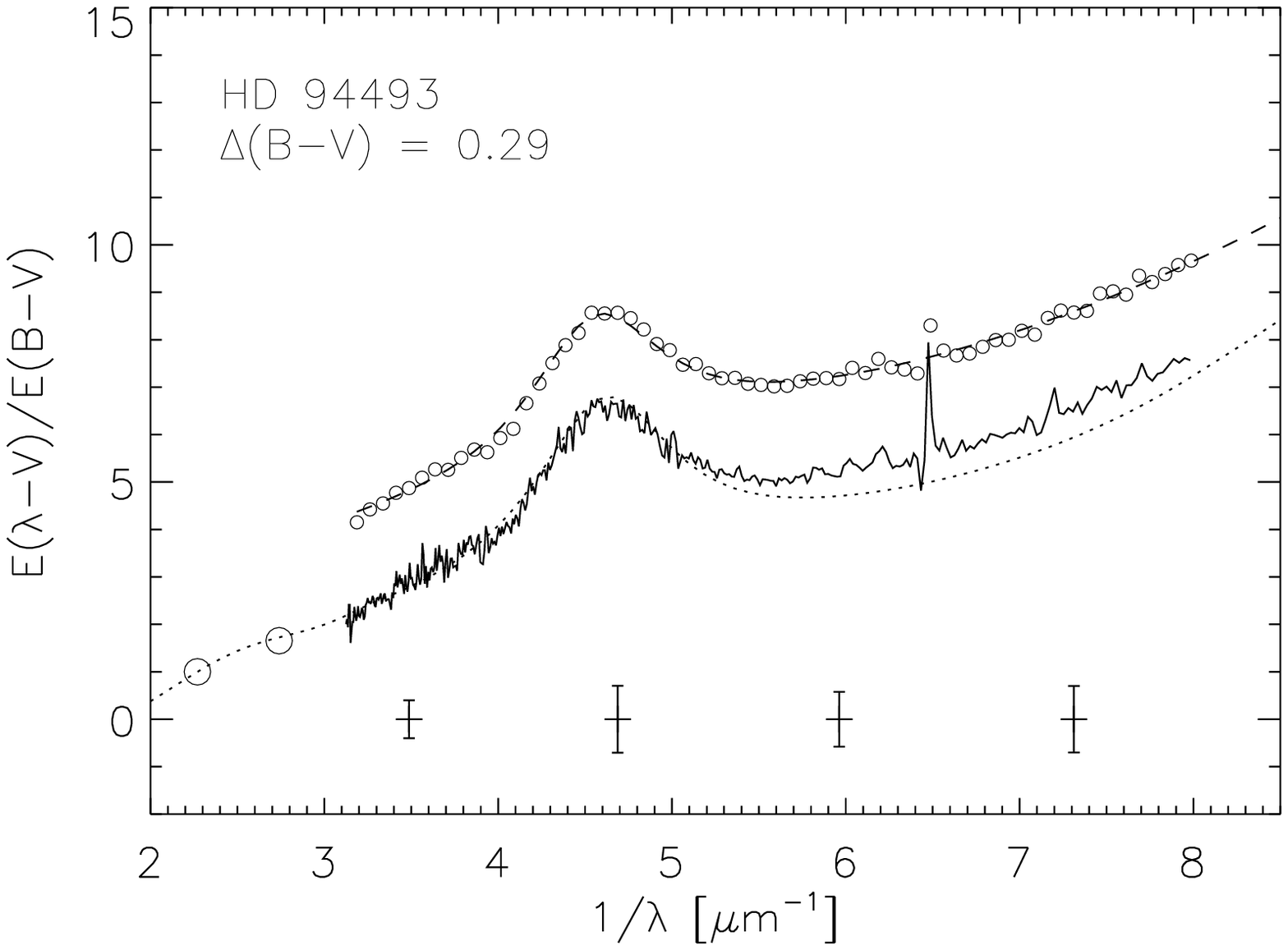} \\
\plottwo{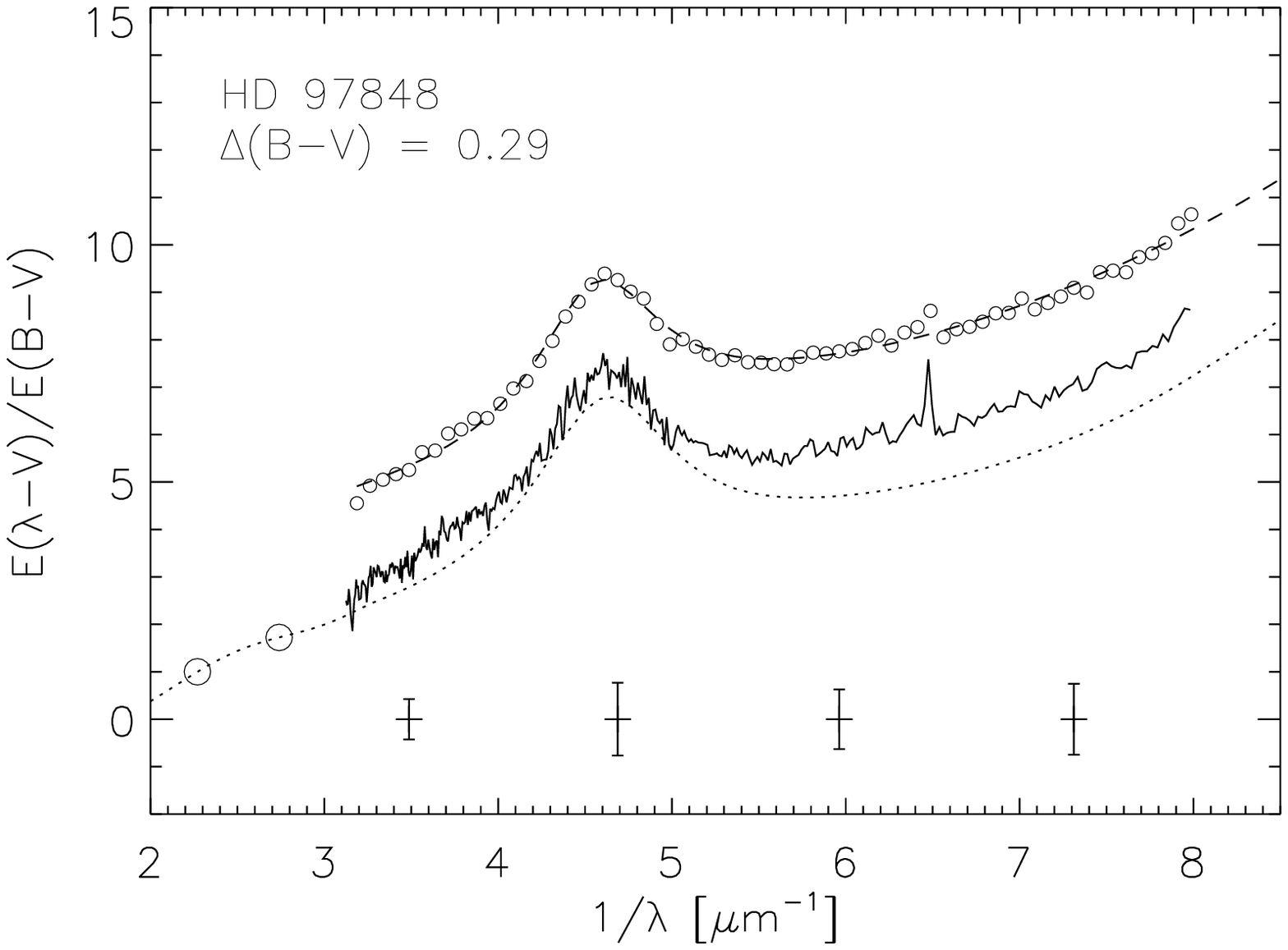}{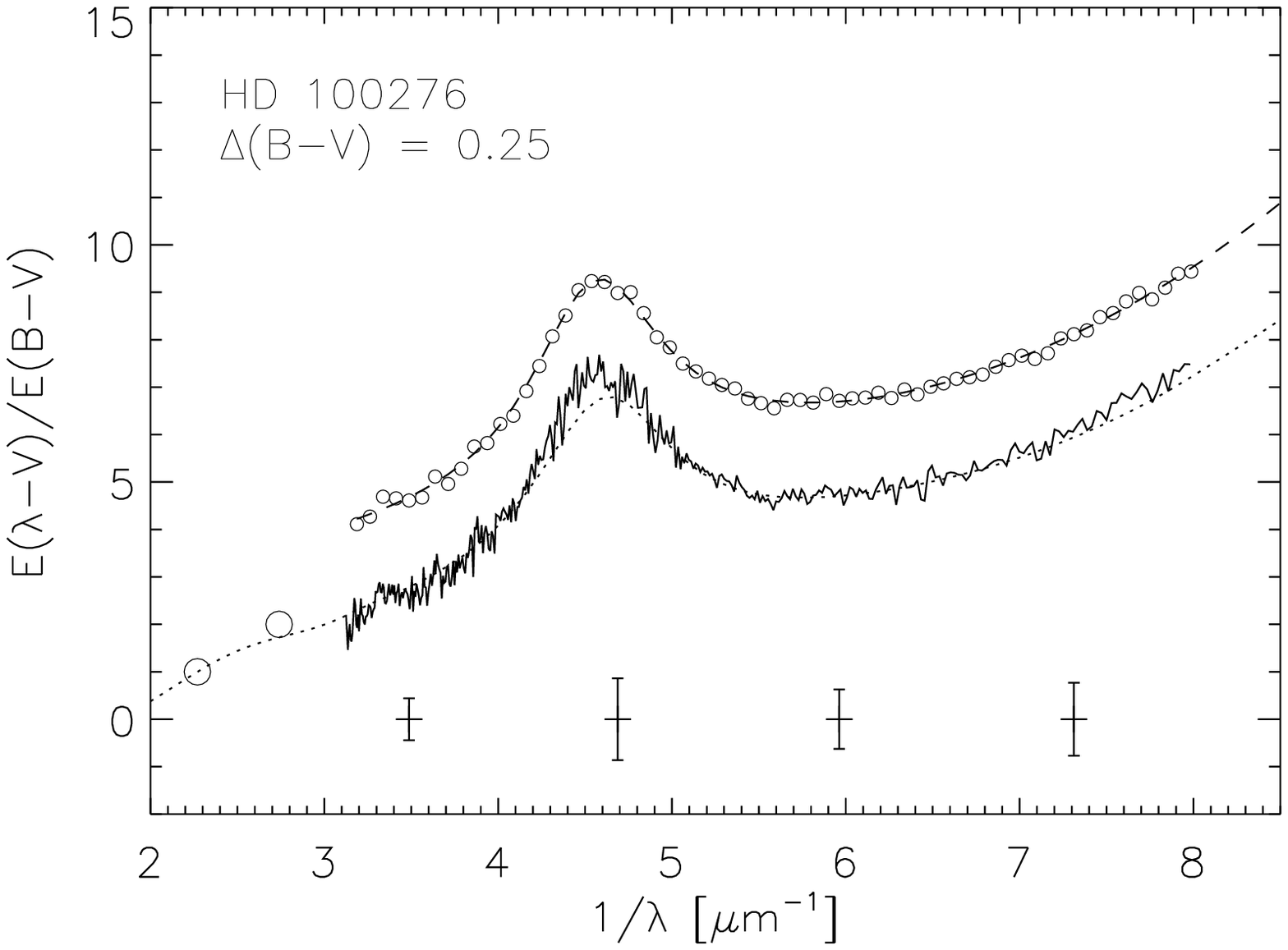} 
\caption{Extinction curves for the stars in the sample. Each panel
shows the calculated extinction curve (solid line), smoothed curve
(small open circles, offset two units for clarity), FM fit (dashed
line), CCM relation for $R_V$=3.1 (dotted line), and extinction in the
U and B bands (large open circles).  One sigma error bars for the
unsmoothed extinction curve are plotted at four representative
wavelengths. The value of $\Delta (B-V)$ between the reddened and
comparison stars is listed.}
\end{figure*}

\begin{figure*}
\figurenum{2 (cont.)}
\plottwo{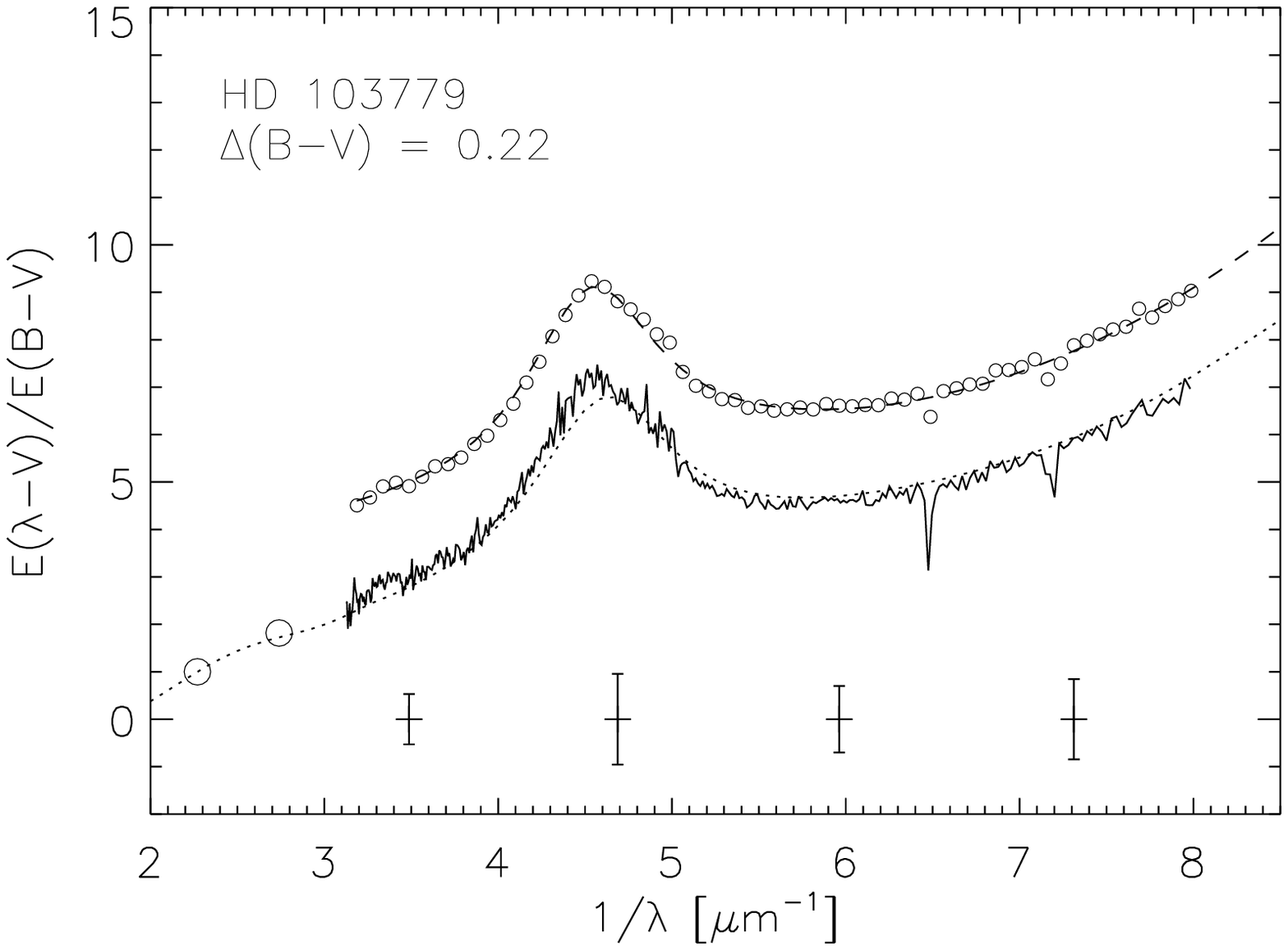}{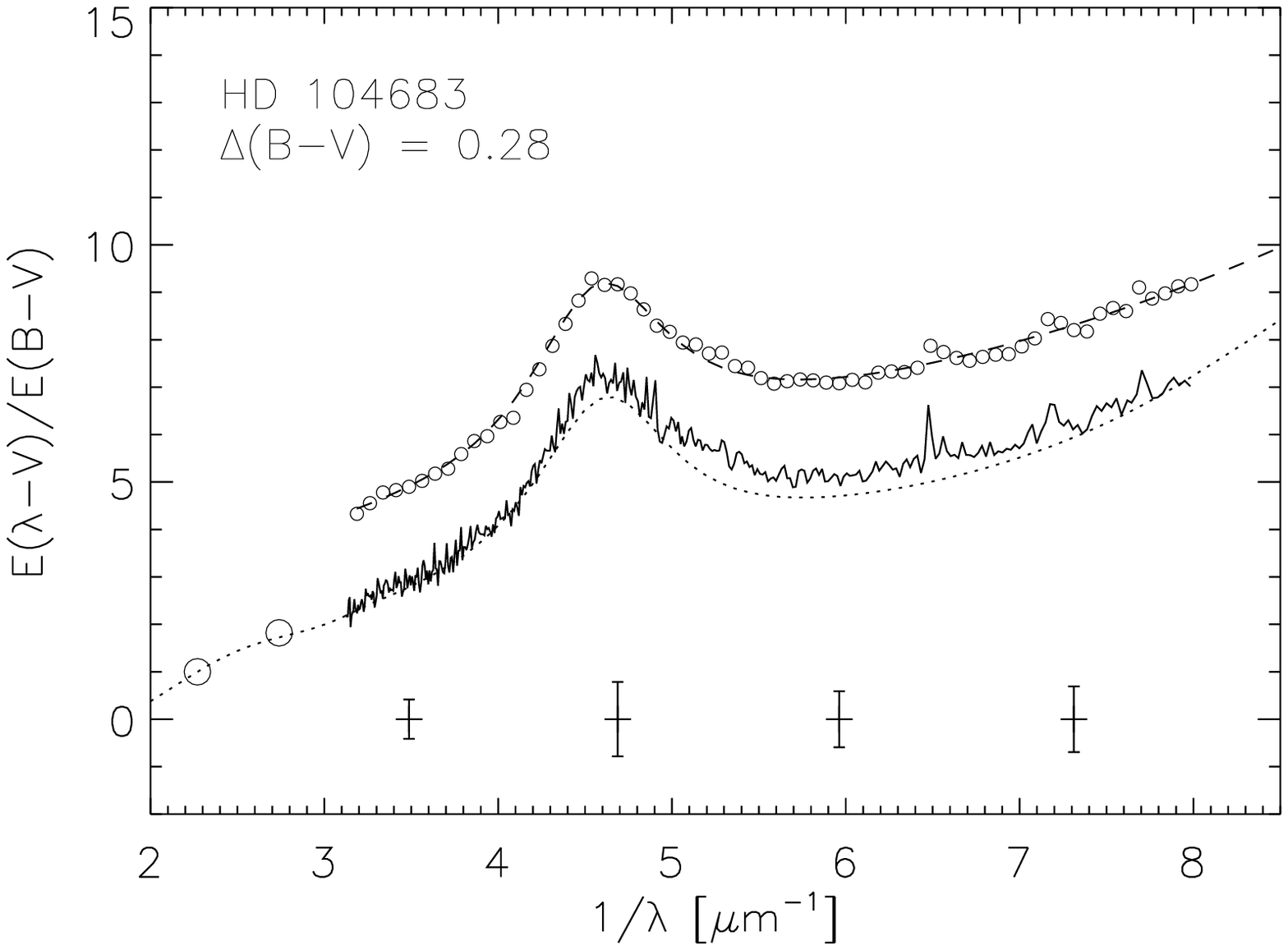}\\
\plottwo{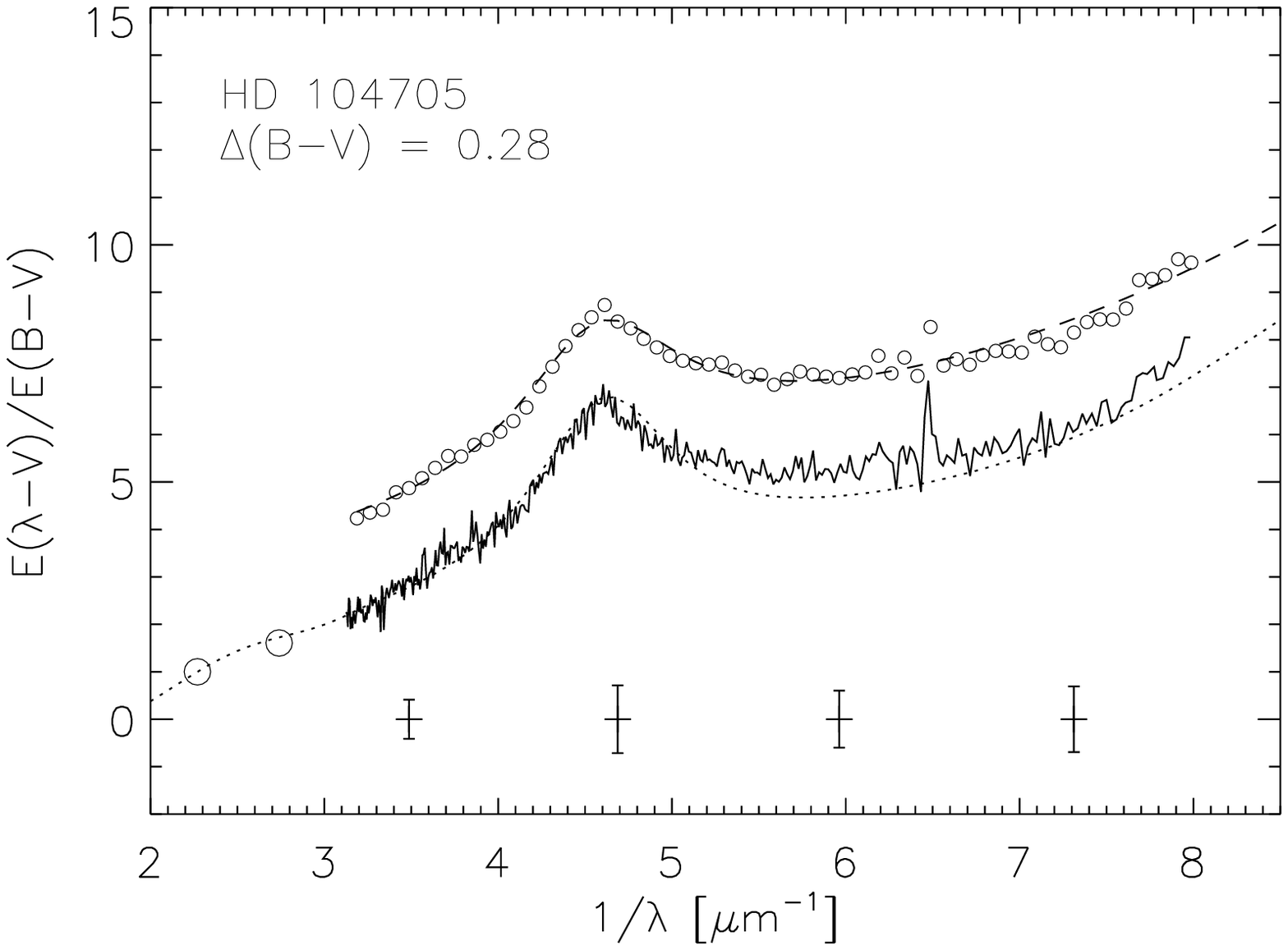}{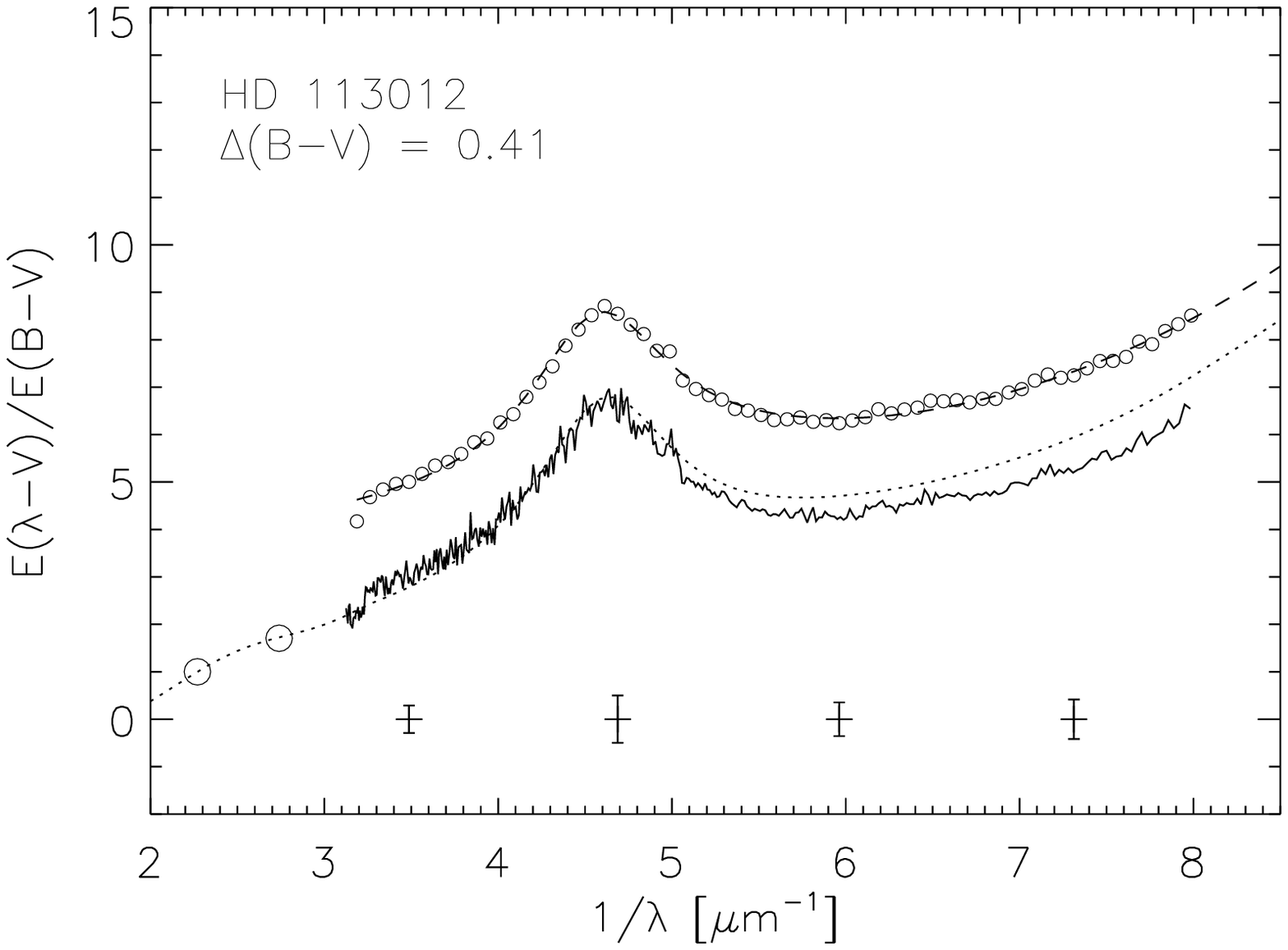} \\
\plottwo{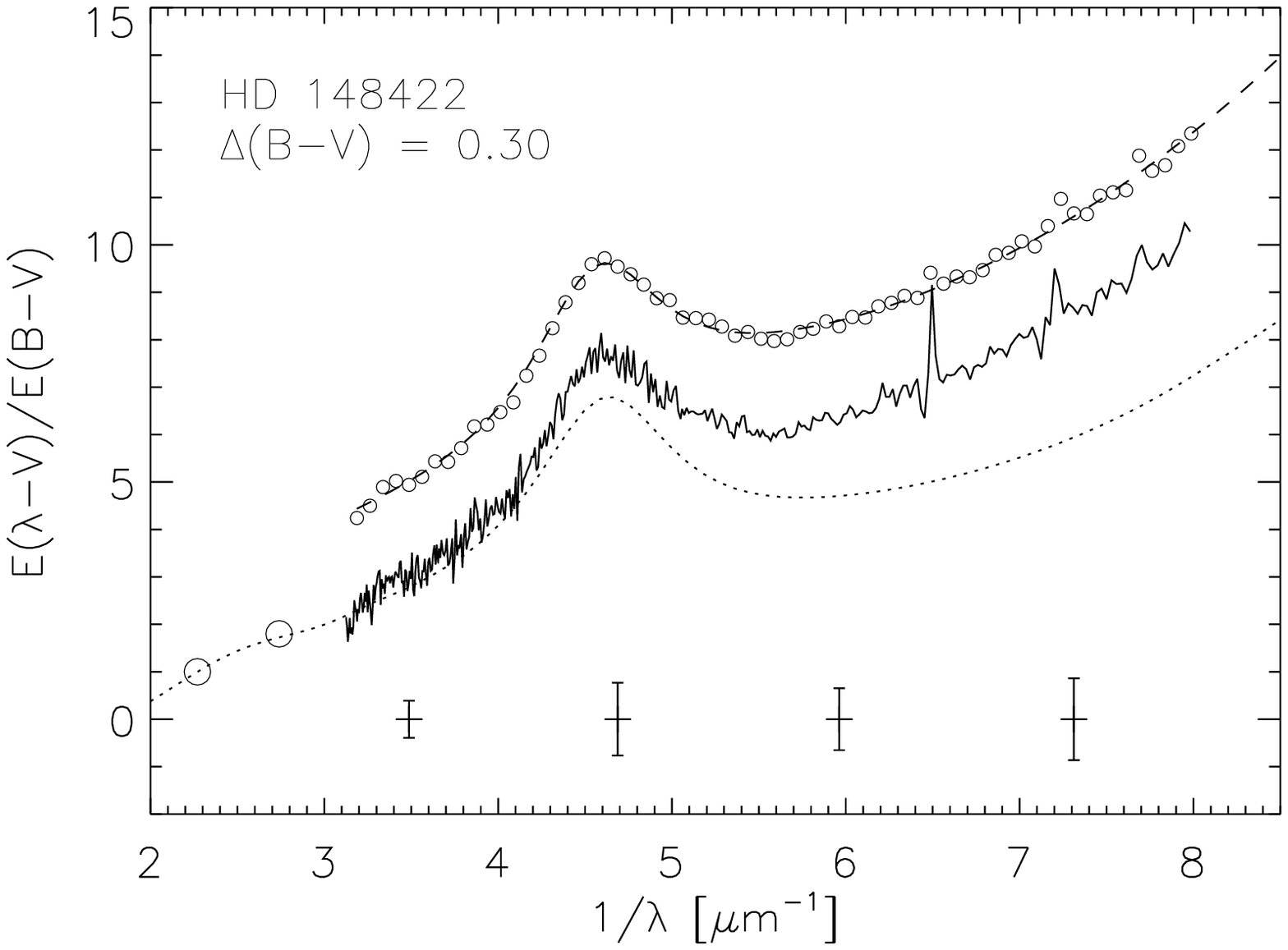}{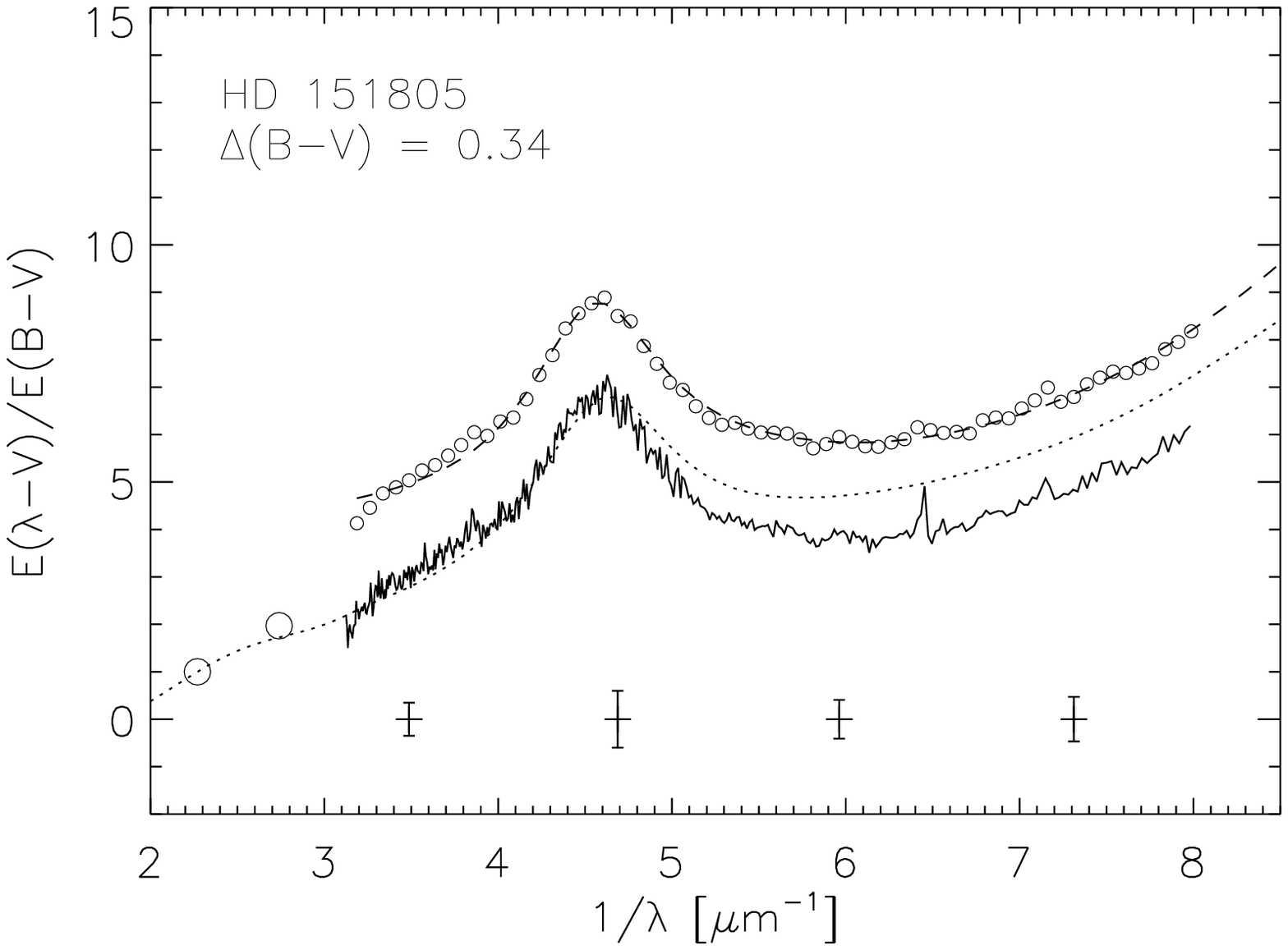} 
\caption{Extinction curves for the stars in the sample.}
\end{figure*}

\begin{figure*}
\figurenum{2 (cont.)}
\plottwo{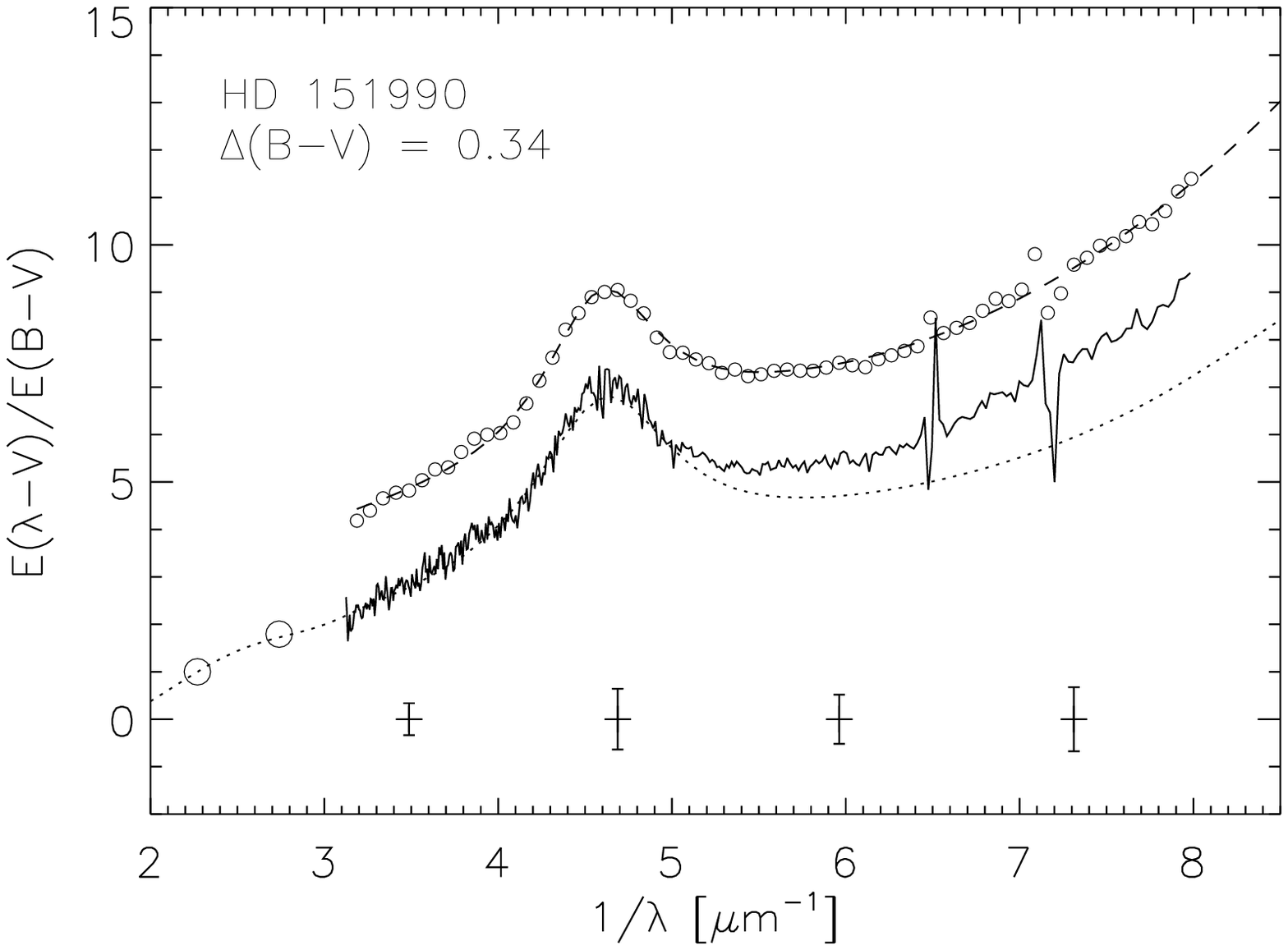}{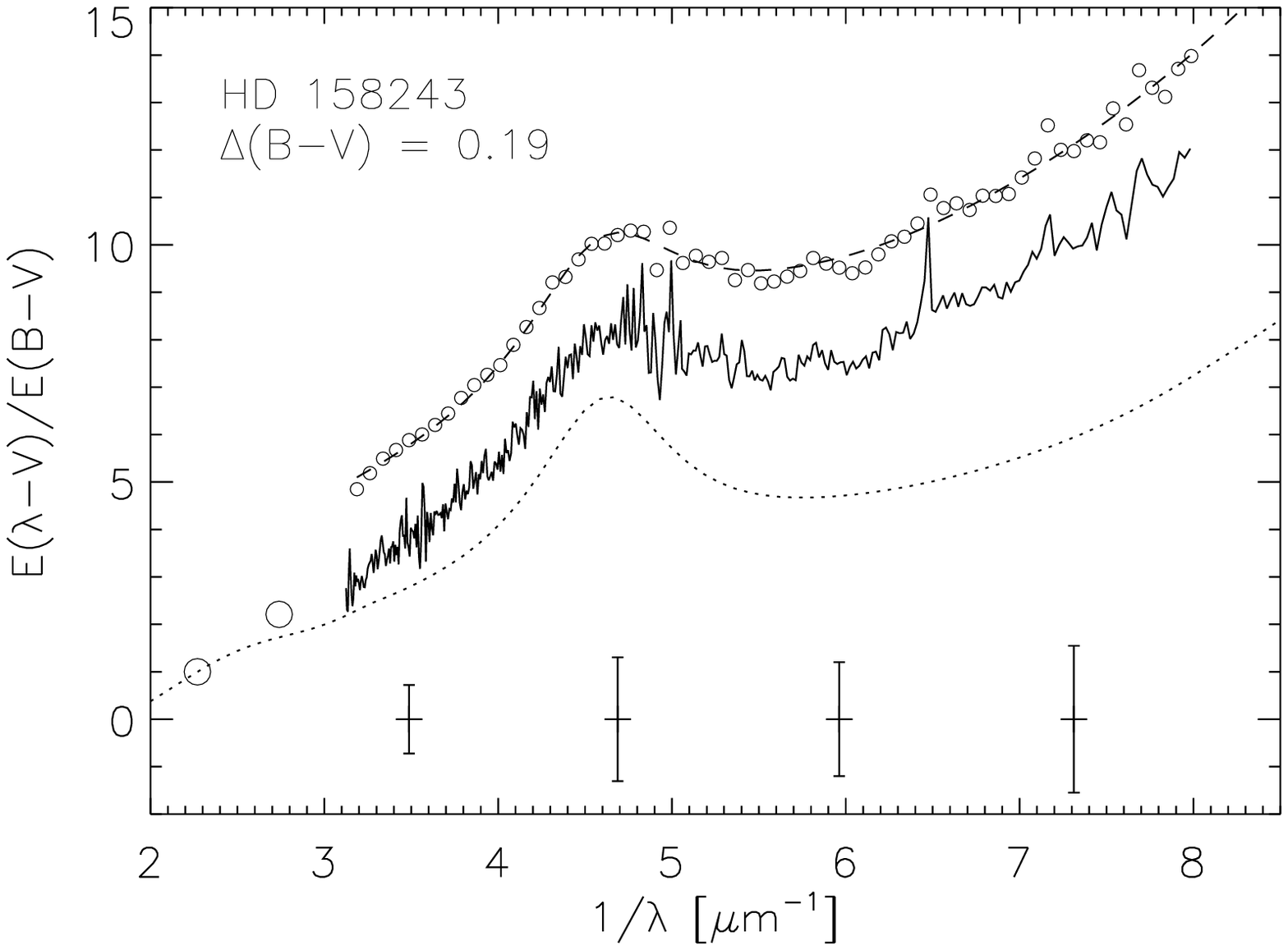} \\
\plottwo{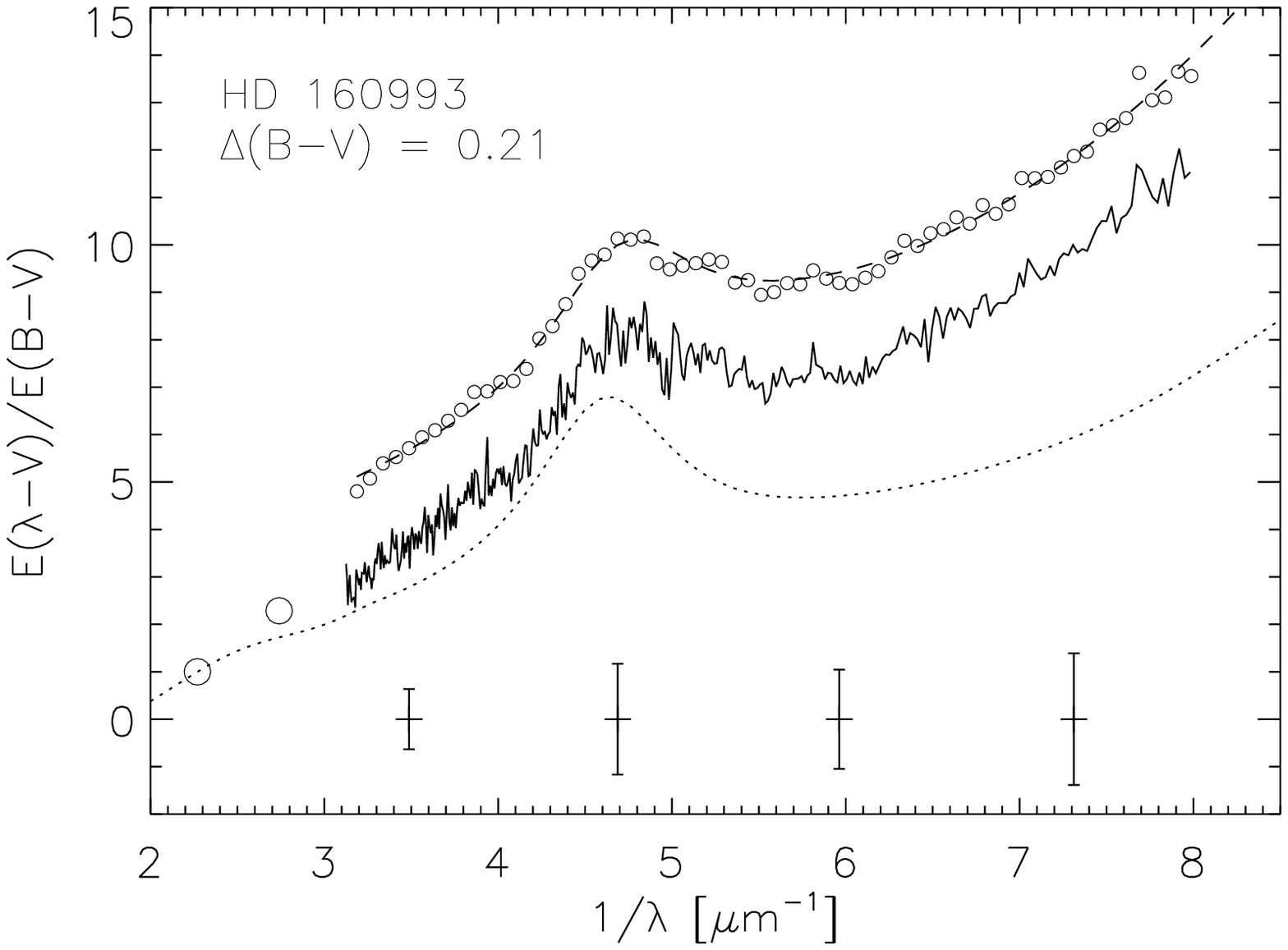}{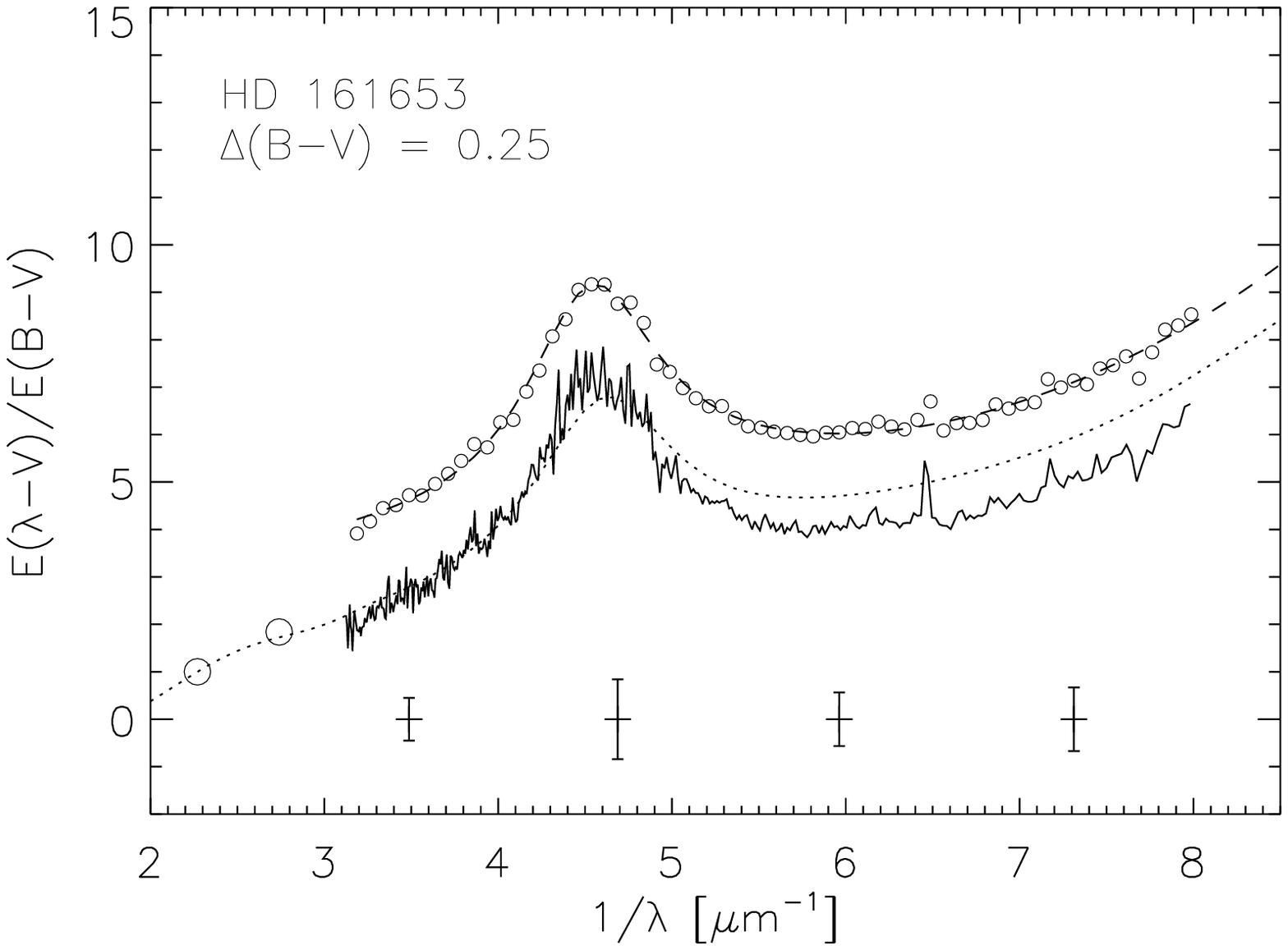}\\
\plottwo{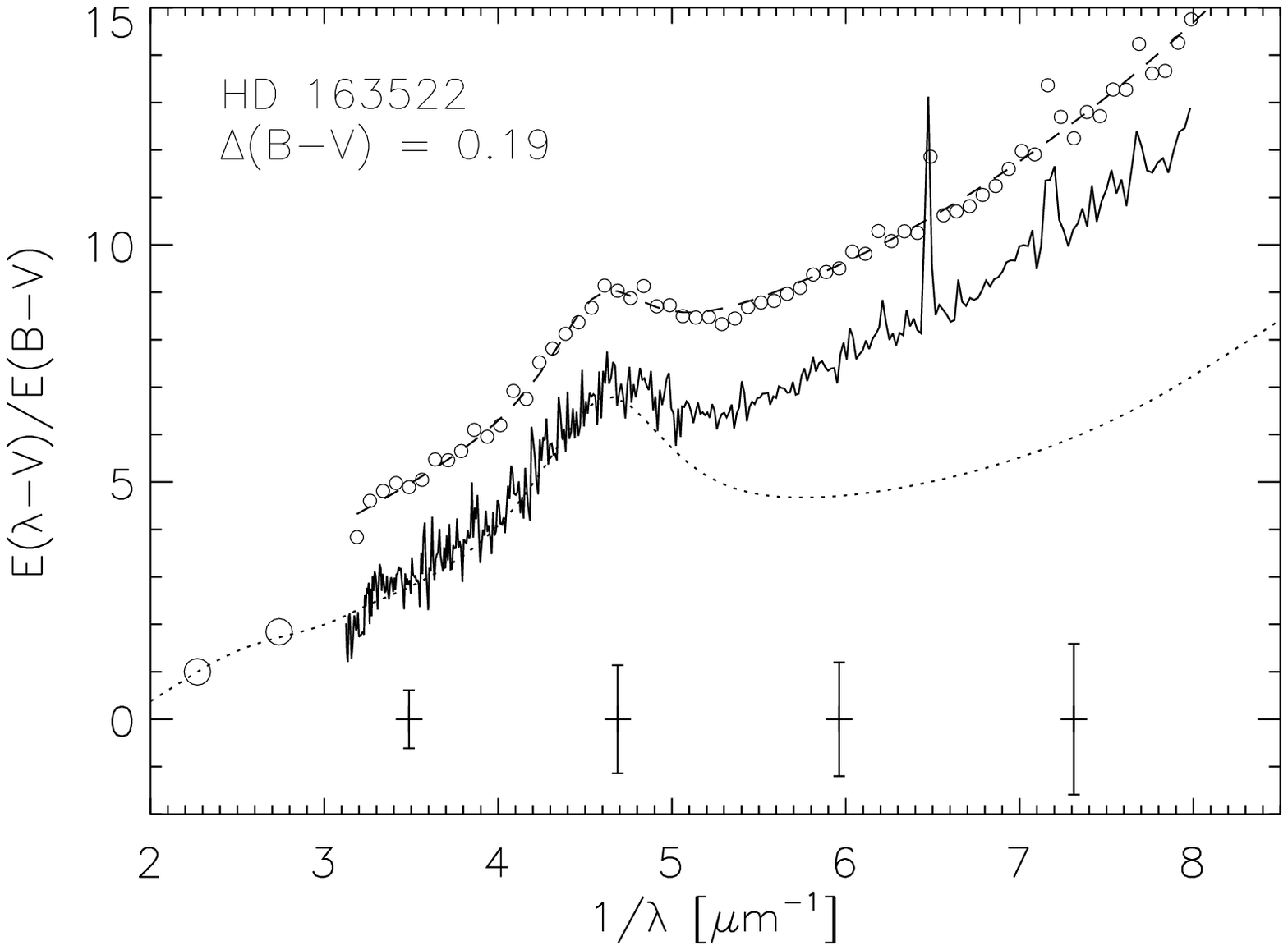}{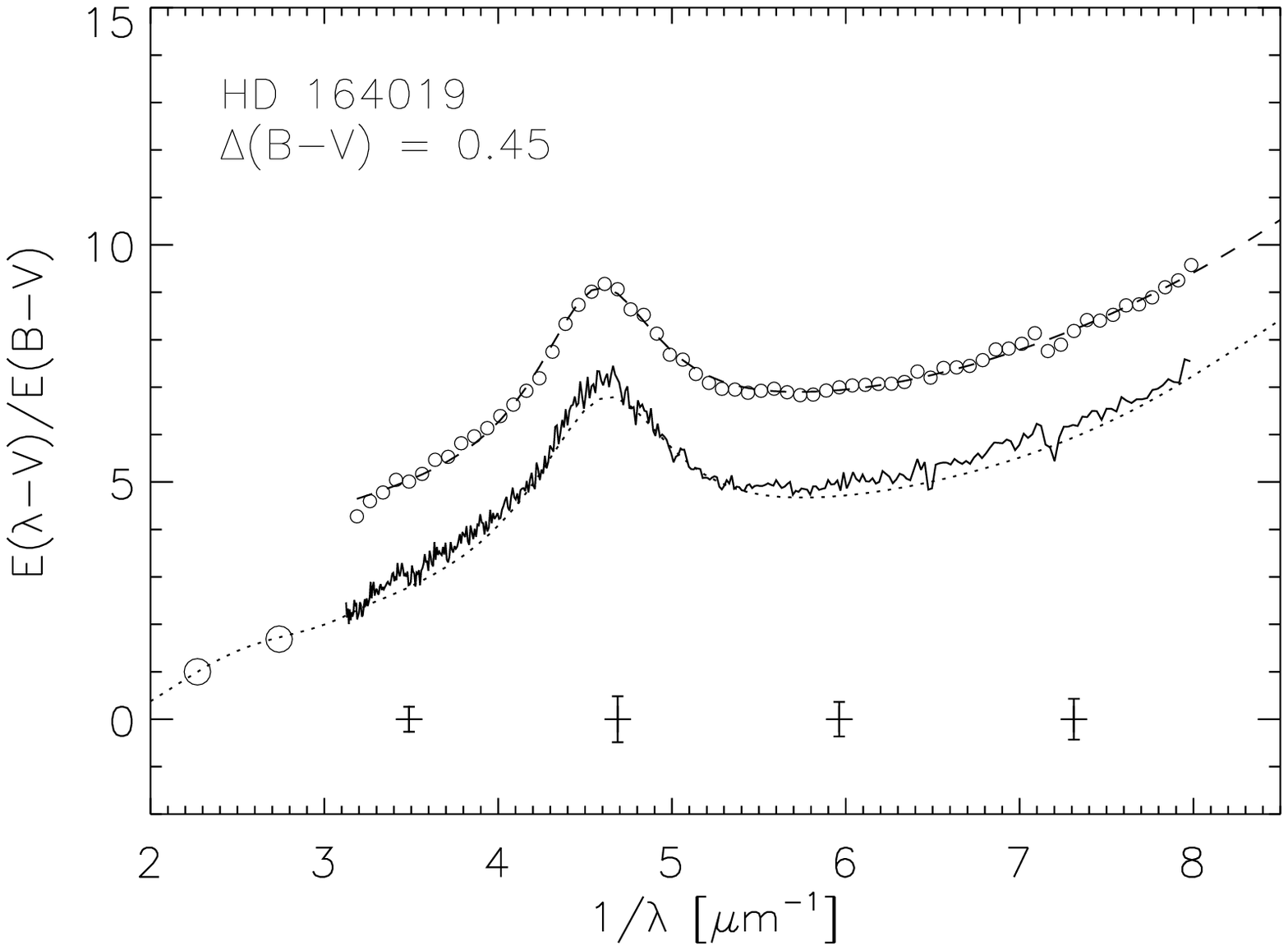} 
\caption{Extinction curves for the stars in the sample.}
\end{figure*}

\begin{figure*}
\figurenum{2 (cont.)}
\plottwo{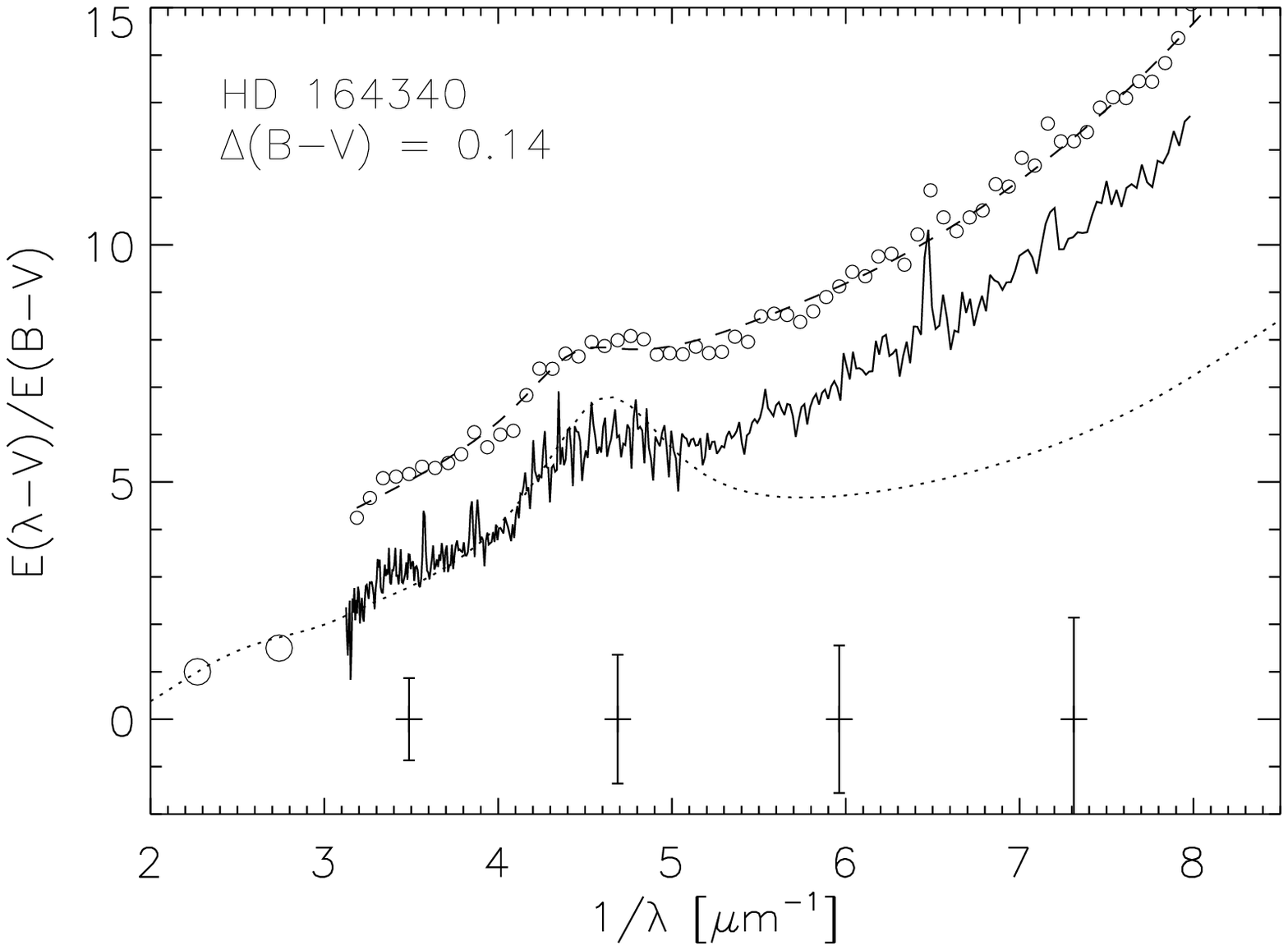}{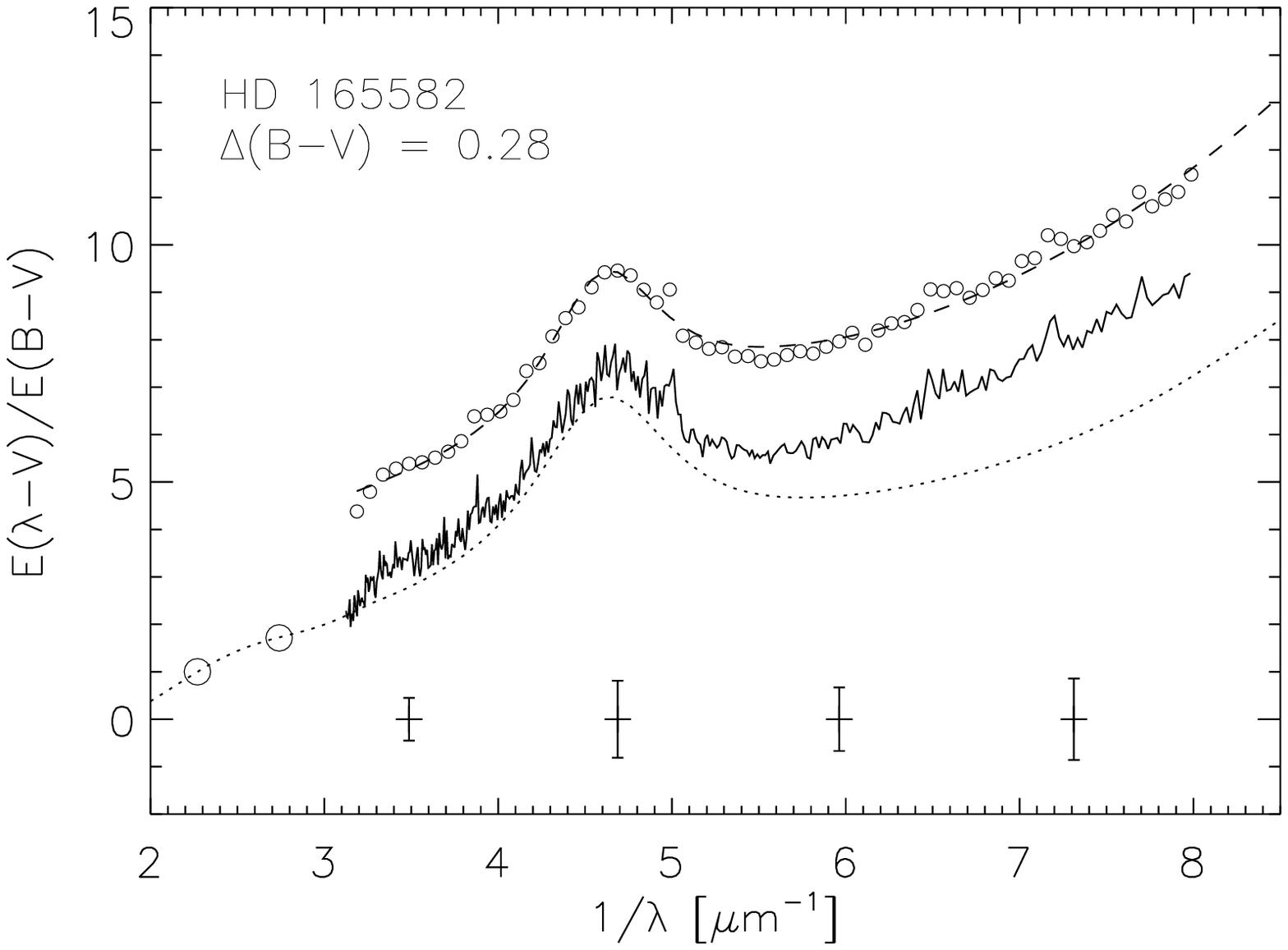} \\
\plottwo{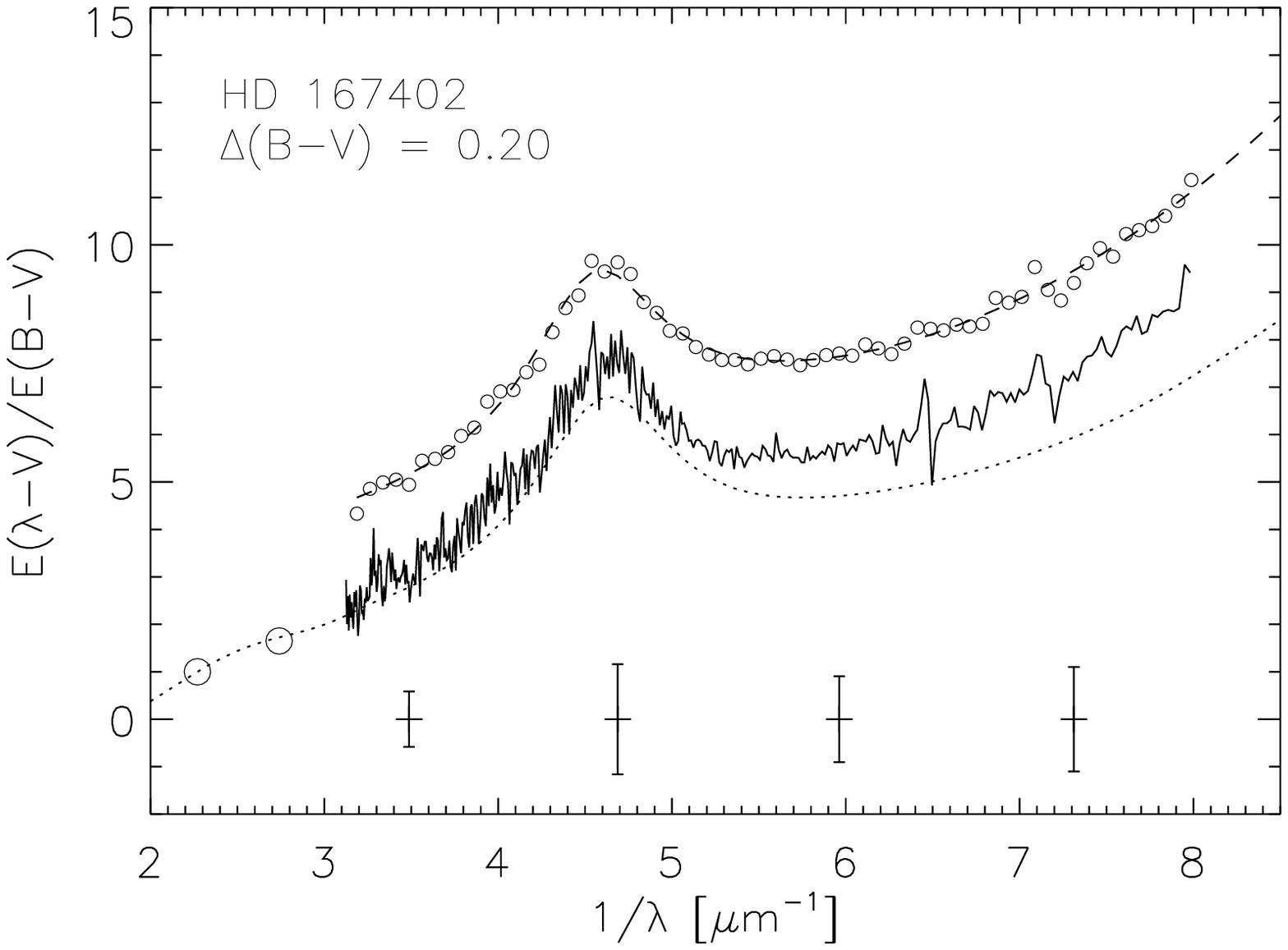}{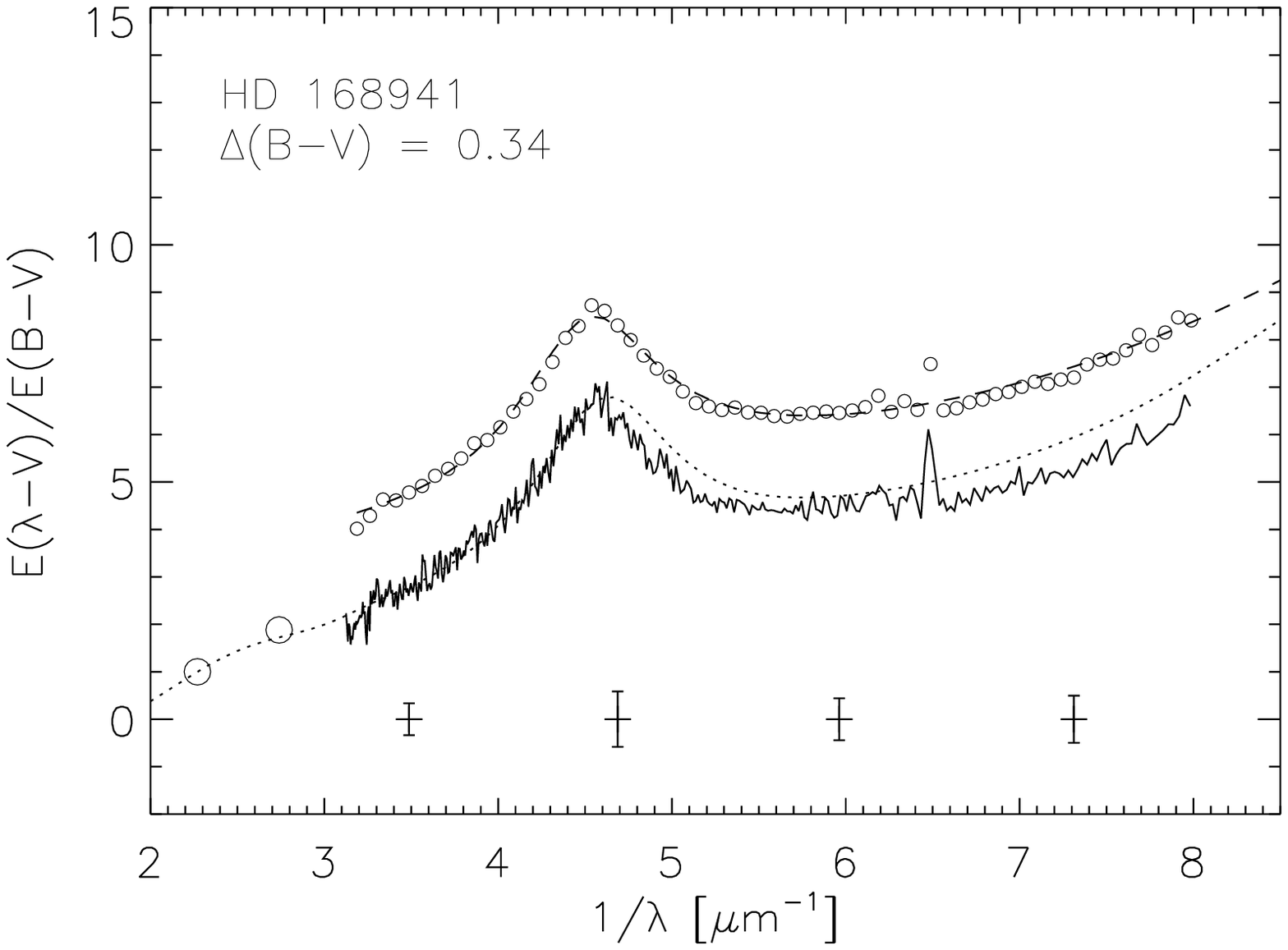} \\
\plottwo{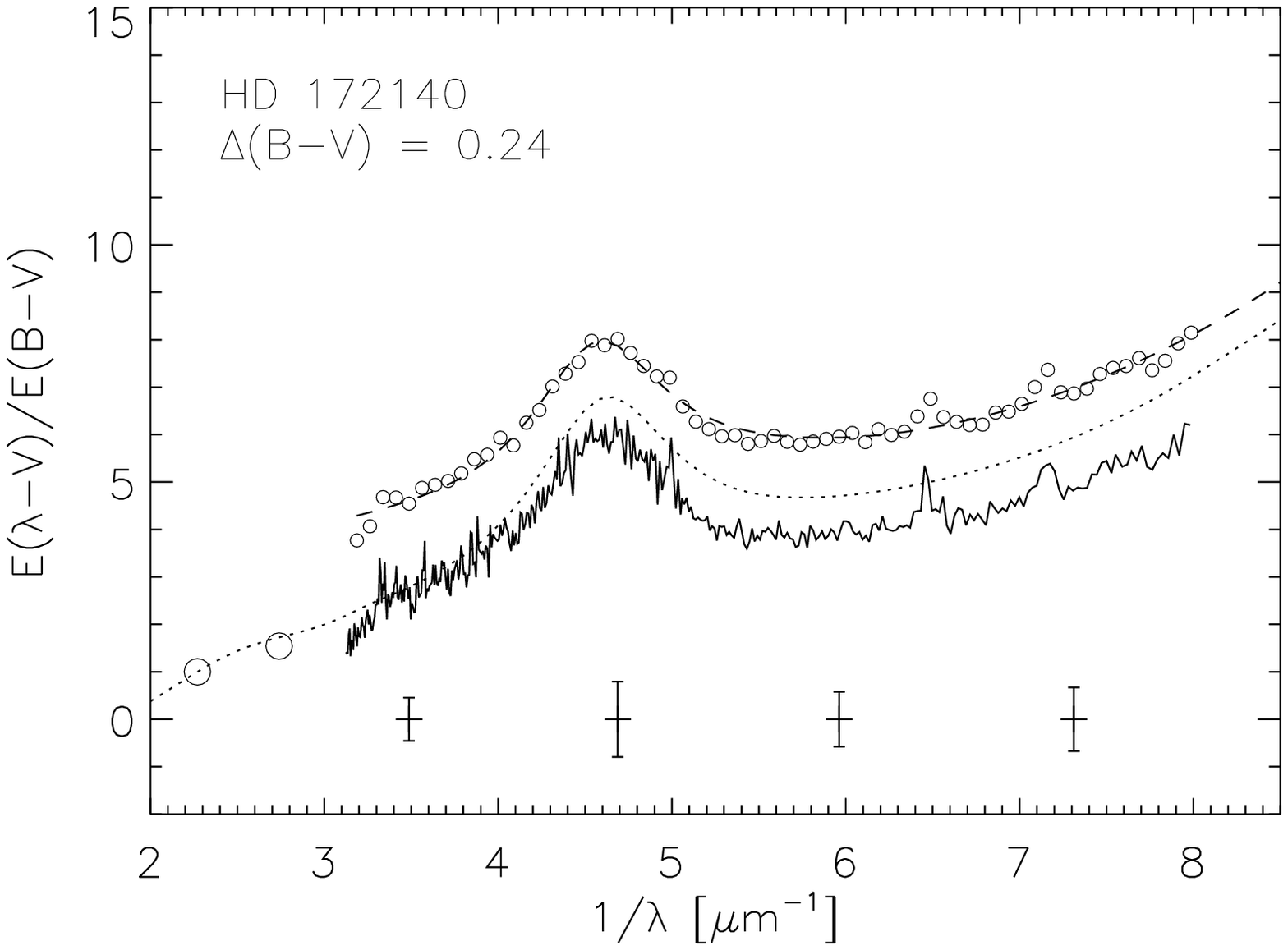}{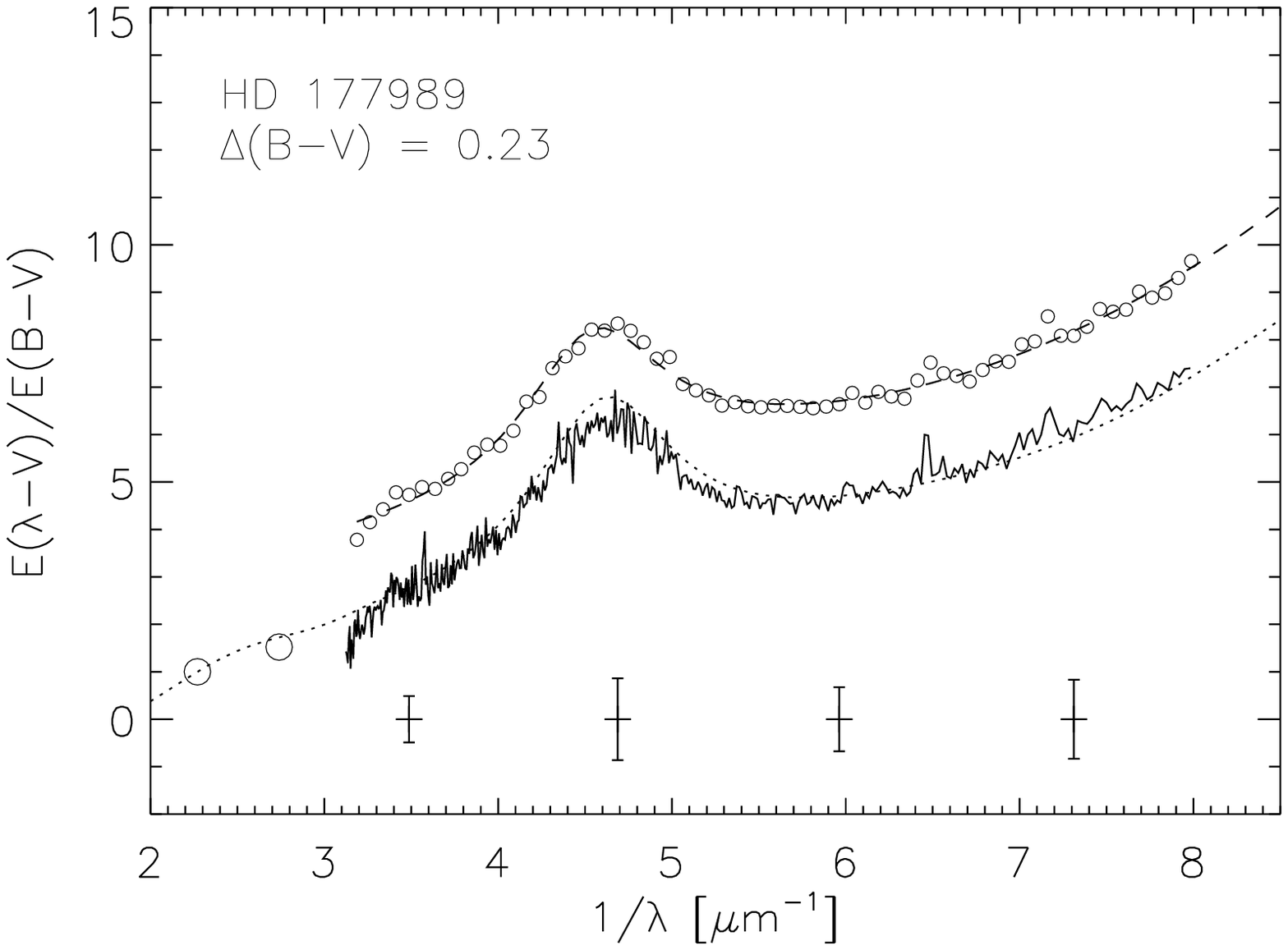}\\
\plottwo{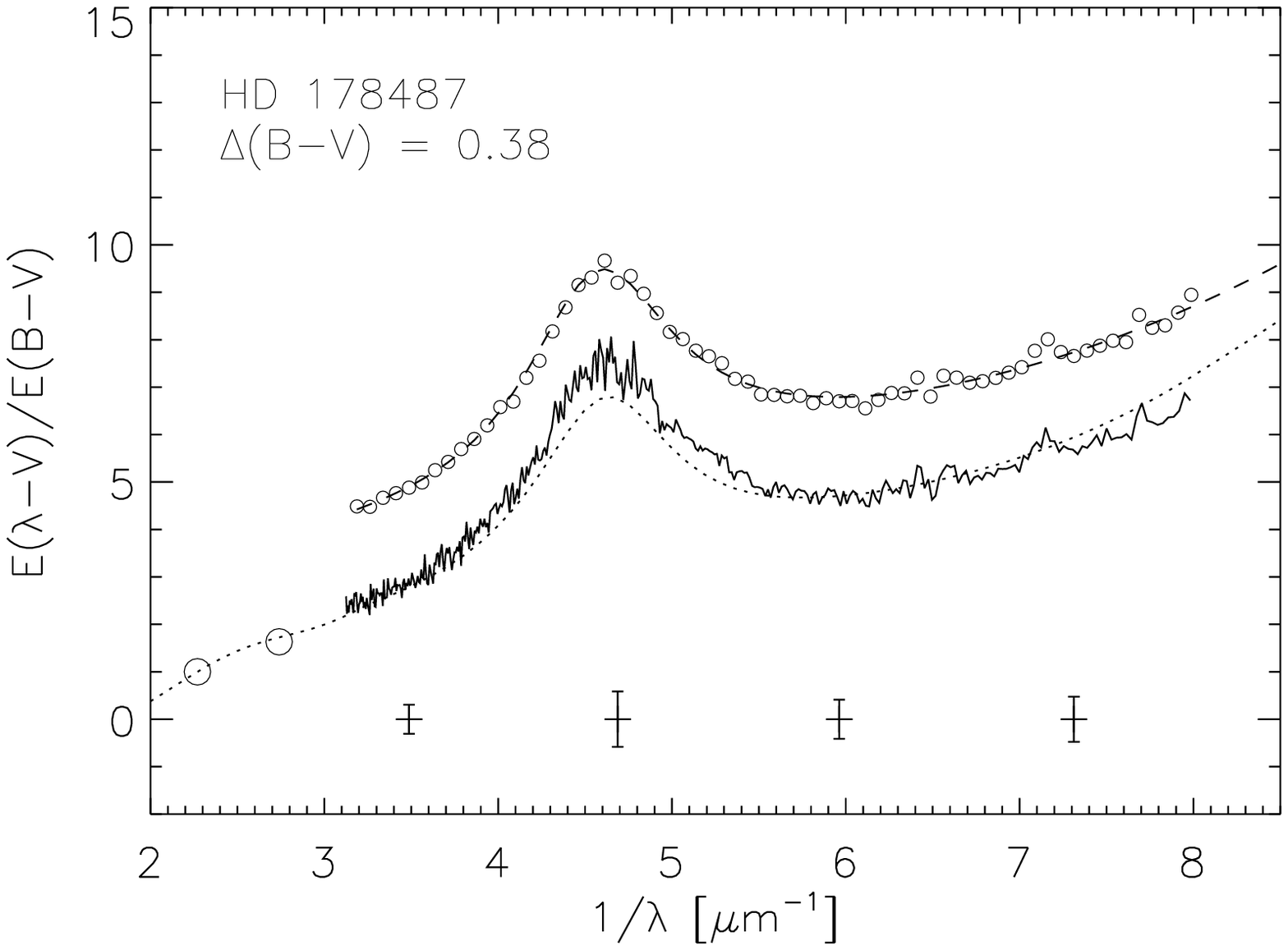}{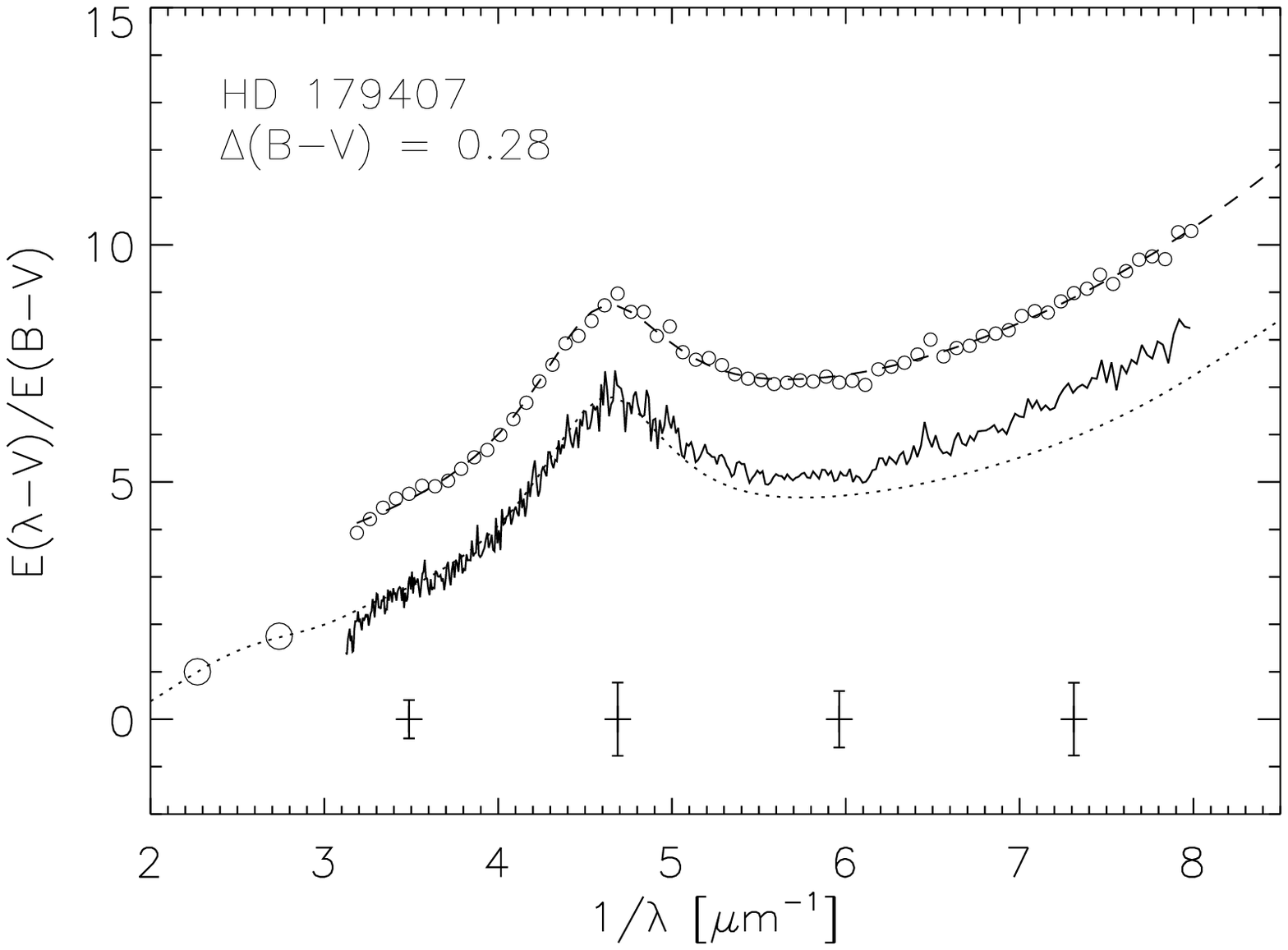}
\caption{Extinction curves for the stars in the sample.}
\end{figure*}

\begin{deluxetable}{lcccccc}
\tablewidth{0pt}
\small
\tablecaption{FM Parameters}
\tablehead{\colhead{Star} & \colhead{$c_1$} & \colhead{$c_2$} & 
           \colhead{$c_3$} & \colhead{$c_4$} &
           \colhead{$x_o$} & \colhead{$\gamma$} }
\startdata
\multicolumn{7}{c}{Low Density Sample} \nl
\tableline
  HD 64219 & $-0.83 \pm  0.22$ & $ 0.98 \pm  0.10$ & $ 4.36 \pm  0.33$ & $ 0.93 \pm  0.14$ & $ 4.61 \pm  0.01$ & $ 1.01 \pm  0.03$ \nl
  HD 69106 & $ 0.40 \pm  0.29$ & $ 0.58 \pm  0.06$ & $ 3.40 \pm  0.26$ & $ 0.55 \pm  0.06$ & $ 4.58 \pm  0.01$ & $ 1.01 \pm  0.03$ \nl
  HD 93827 & $-0.54 \pm  0.15$ & $ 0.83 \pm  0.07$ & $ 4.18 \pm  0.14$ & $ 0.40 \pm  0.03$ & $ 4.59 \pm  0.01$ & $ 1.07 \pm  0.03$ \nl
  HD 94493 & $-0.95 \pm  0.08$ & $ 0.97 \pm  0.08$ & $ 2.75 \pm  0.17$ & $ 0.26 \pm  0.02$ & $ 4.57 \pm  0.01$ & $ 0.95 \pm  0.01$ \nl
  HD 97848 & $-0.32 \pm  0.17$ & $ 0.95 \pm  0.07$ & $ 2.57 \pm  0.15$ & $ 0.34 \pm  0.04$ & $ 4.58 \pm  0.01$ & $ 0.89 \pm  0.01$ \nl
 HD 100276 & $-0.66 \pm  0.18$ & $ 0.81 \pm  0.07$ & $ 3.69 \pm  0.20$ & $ 0.55 \pm  0.07$ & $ 4.57 \pm  0.01$ & $ 0.93 \pm  0.02$ \nl
 HD 103779 & $ 0.31 \pm  0.18$ & $ 0.63 \pm  0.08$ & $ 3.46 \pm  0.37$ & $ 0.57 \pm  0.04$ & $ 4.54 \pm  0.02$ & $ 0.93 \pm  0.00$ \nl
 HD 104683 & $-0.75 \pm  0.12$ & $ 0.91 \pm  0.08$ & $ 3.62 \pm  0.14$ & $ 0.19 \pm  0.02$ & $ 4.59 \pm  0.01$ & $ 0.98 \pm  0.02$ \nl
 HD 104705 & $-0.86 \pm  0.14$ & $ 0.92 \pm  0.07$ & $ 3.89 \pm  0.15$ & $ 0.30 \pm  0.03$ & $ 4.59 \pm  0.01$ & $ 1.13 \pm  0.03$ \nl
 HD 113012 & $ 0.64 \pm  0.18$ & $ 0.54 \pm  0.03$ & $ 3.51 \pm  0.11$ & $ 0.49 \pm  0.03$ & $ 4.59 \pm  0.01$ & $ 1.00 \pm  0.02$ \nl
 HD 151805 & $ 1.26 \pm  0.15$ & $ 0.35 \pm  0.04$ & $ 3.48 \pm  0.27$ & $ 0.71 \pm  0.03$ & $ 4.57 \pm  0.01$ & $ 0.94 \pm  0.01$ \nl
 HD 161653 & $ 0.05 \pm  0.27$ & $ 0.57 \pm  0.04$ & $ 3.93 \pm  0.19$ & $ 0.55 \pm  0.06$ & $ 4.55 \pm  0.01$ & $ 0.93 \pm  0.02$ \nl
 HD 164019 & $-0.00 \pm  0.13$ & $ 0.76 \pm  0.04$ & $ 2.84 \pm  0.08$ & $ 0.43 \pm  0.03$ & $ 4.58 \pm  0.01$ & $ 0.88 \pm  0.01$ \nl
 HD 167402 & $-0.78 \pm  0.20$ & $ 1.00 \pm  0.11$ & $ 3.28 \pm  0.20$ & $ 0.63 \pm  0.10$ & $ 4.56 \pm  0.01$ & $ 0.94 \pm  0.03$ \nl
 HD 168941 & $-0.12 \pm  0.18$ & $ 0.69 \pm  0.04$ & $ 3.35 \pm  0.11$ & $ 0.32 \pm  0.03$ & $ 4.53 \pm  0.01$ & $ 0.98 \pm  0.02$ \nl
 HD 172140 & $ 0.39 \pm  0.29$ & $ 0.53 \pm  0.04$ & $ 2.80 \pm  0.15$ & $ 0.49 \pm  0.06$ & $ 4.58 \pm  0.01$ & $ 0.94 \pm  0.02$ \nl
 HD 177989 & $-0.81 \pm  0.21$ & $ 0.85 \pm  0.07$ & $ 3.10 \pm  0.20$ & $ 0.49 \pm  0.07$ & $ 4.56 \pm  0.01$ & $ 0.99 \pm  0.02$ \nl
 HD 178487 & $-0.32 \pm  0.15$ & $ 0.74 \pm  0.04$ & $ 4.78 \pm  0.19$ & $ 0.32 \pm  0.03$ & $ 4.58 \pm  0.01$ & $ 1.04 \pm  0.01$ \nl
 HD 179407 & $-1.42 \pm  0.13$ & $ 1.03 \pm  0.07$ & $ 3.59 \pm  0.19$ & $ 0.49 \pm  0.06$ & $ 4.60 \pm  0.01$ & $ 1.03 \pm  0.02$ \nl
 Average   & $-0.47$ & $0.66$ & $3.38$ & $0.35$ &  $4.58$ & $0.94$ \nl
\tableline
\multicolumn{7}{c}{Peculiar Velocity Sample (SD region - 2 curves)} \nl
\tableline
 HD 148422 & $-2.13 \pm  0.07$ & $ 1.36 \pm  0.12$ & $ 3.01 \pm  0.23$ & $ 0.53 \pm  0.03$ & $ 4.58 \pm  0.02$ & $ 0.93 \pm  0.00$ \nl
 HD 151990 & $-1.07 \pm  0.09$ & $ 1.04 \pm  0.07$ & $ 2.19 \pm  0.06$ & $ 0.68 \pm  0.06$ & $ 4.60 \pm  0.01$ & $ 0.81 \pm  0.02$ \nl
 HD 158243 & $-2.20 \pm  0.16$ & $ 1.56 \pm  0.22$ & $ 4.26 \pm  0.52$ & $ 0.55 \pm  0.06$ & $ 4.61 \pm  0.02$ & $ 1.16 \pm  0.01$ \nl
 HD 160993 & $-1.65 \pm  0.08$ & $ 1.43 \pm  0.19$ & $ 3.42 \pm  0.31$ & $ 0.74 \pm  0.07$ & $ 4.72 \pm  0.02$ & $ 1.07 \pm  0.01$ \nl
 HD 163522 & $-3.70 \pm  0.20$ & $ 1.86 \pm  0.23$ & $ 1.18 \pm  0.07$ & $ 0.51 \pm  0.10$ & $ 4.59 \pm  0.01$ & $ 0.74 \pm  0.03$ \nl
 HD 164340 & $-2.96 \pm  0.13$ & $ 1.67 \pm  0.28$ & $ 0.79 \pm  0.06$ & $ 0.76 \pm  0.18$ & $ 4.44 \pm  0.02$ & $ 0.79 \pm  0.03$ \nl
 HD 165582 & $-0.87 \pm  0.12$ & $ 1.10 \pm  0.09$ & $ 2.31 \pm  0.14$ & $ 0.57 \pm  0.06$ & $ 4.61 \pm  0.01$ & $ 0.84 \pm  0.01$ \nl
 Average & $-1.83$ & $1.19$ & $1.52$ & $0.58$ & $4.60$ & $0.93$ \nl
\enddata
\end{deluxetable}

Extinction curves were constructed using the standard pair method
(e.g., Massa, Savage \& Fitzpatrick 1983).  Uncertainties in the
extinction curves contain terms that depend both on the broadband
photometric uncertainties as well as those in the IUE fluxes, which
are calculated directly in NEWSIPS.  Our error analysis is described
in detail in Gordon \& Clayton (1998).  The sample includes early type
supergiants which may be used with the same accuracy as main sequence
stars in calculating extinction (Cardelli, Sembach, \& Mathis 1992).
We required $\Delta (B-V) \ge 0.14$ between the reddened and
comparison stars to minimize the uncertainties.  The comparison stars
have been dereddened as described by Cardelli et al.  (1992).  Table 1
lists a value of E(B-V) for each star from Sembach et
al. (1993). Table 2 lists the $\Delta (B-V)$ between the measured
(B-V) of the reddened star and the $(B-V)_o$ of the best-match
dereddened comparison star.  The extinction curves for the sample
stars are shown in Figure 2.

The extinction curves have been fitted using the Fitzpatrick \& Massa
(1990, hereafter FM) parameterization.  They have developed an
analytical representation of the shape of the extinction curves using
a small number of parameters.  This was done using linear combinations
of a Drude bump profile, $D(x;\gamma,x_o)$, a linear background and a
far-UV curvature function, $F(x)$, where $x = \lambda^{-1}$.  There
are 6 parameters determined in the fit: The strength, central
wavelength, and width of the bump, $c_3$, $x_o$, and $\gamma$, the
slope and intercept of the linear background, $c_1$ and $c_2$, and the
strength of the far-UV curvature, $c_4$.  The FM fits to individual
extinction curves are plotted in Figure 2 and the best fit parameters
for each curve are given in Table 3.

Near-infrared photometry exists for a few of the reddened stars in our
sample.  For these stars, using JHK photometry, we calculated values
of R$_V$.  Due the small values of E(B-V) for our sample stars, the
uncertainties in R$_V$ are relatively large.  Within these
uncertainties, most are consistent with the typical diffuse dust value
of 3.1.  There is no trend with position on the sky discernible with
the small number of sightlines having measured $R_V$ values.

\section{Discussion}

The sightlines in our sample cover very long distances and have
relatively low reddenings so the average densities are amongst the
lowest known (Sembach et al.  1993).  The measured values for
$n_o$(H~I) (= N(H~I)sin$\vert b \vert$/h(H~I) where h(H~I) is the
scale height of H~I) are listed in Table 1. The parameter, $n_o$(H~I),
is a measure of average density along a sightline (Sembach et
al. 1993).  Typically, when $n_o$(H I) $<$ 0.42 $cm^{-3}$, no large
cold clouds are present along the line of sight (Sembach et al.
1993).  All the stars in Table 1 satisfy this criterion.  In addition,
the warm intercloud medium dominates over the diffuse cold cloud
medium if $n_o$(H I) $<$ 0.2 $cm^{-3}$.  Most of the stars in Table 1
satisfy this criterion or come close to it.  The ratio of the column
density of Ca II to Na I, also listed in Table 1, is another measure
of the relative contributions of cloud and intercloud medium.  Na I is
relatively stronger in clouds while Ca II is relatively strong in the
diffuse ISM.  This is due to the strong variation in the calcium
depletion from the gas phase into dust grains (Sembach \& Danks 1994;
Crinklaw, Federman, \& Joseph 1994).  The depletion is higher inside
clouds and lower outside where the harsher environment including
sputtering and grain collisions will return calcium to the gas phase.
The wide range in N(Ca II)/N(Na I) indicates that the
cloud/inter-cloud fraction varies strongly from one sightline to
another.  The absorption features along these lines of sight show
multiple components indicating that the distribution of gas is patchy.
The average number of components or clouds is 1.5 kpc$^{-1}$ for Na I
and 2.0 kpc$^{-1}$ for Ca II (Sembach \& Danks 1994).  Using these
values and the measured reddening, the average E(B-V) per cloud is
0.05 mag which is typical for standard diffuse clouds (Spitzer 1978).
Similarly, the average H I column density per cloud is 2.3 x 10$^{20}$
cm $^{-2}$.  It is likely that most of these sightlines are dominated
by warm intercloud medium and have little contribution from the cold
cloud medium.

\vspace*{0.05in}
\figurenum{3}
\begin{center}
\plotfiddle{f3.eps}{2.5in}{0}{70}{70}{-220}{-130}
\end{center}
\figcaption{The steepness of the far-UV extinction ($c_2$) plotted
against bump strength ($\pi c_3/ 2 \gamma$) for the stars in the SD
region lying between l=325$^o$ to 0$^o$ (filled circles) and those of
the rest of the sample (open circles). Also, plotted from upper left
to lower right are average points for the Galaxy, the LMC (outside LMC
2), LMC 2 and SMC (Misselt et al. 1999; Gordon \& Clayton 1998)
(filled squares). }
\vspace*{0.2in}

Figure 2 shows that even for extremely diffuse sightlines such as
those in our sample, most extinction curves still follow CCM with
$R_V$ = 3.1.  However, there is a subsample of these sightlines whose
extinction curves show weak bumps and very steep far-UV extinction
similar to the Magellanic clouds.  These sightlines all lie in one
region of the sky in the direction of the Galactic center.  This
region (Sembach \& Danks 1994 (hereafter the SD region)) coincides
with an area of Galactic longitude, l=325$^o$ to 0$^o$, where large
forbidden velocities have been observed in the gas.  Figure 3 shows
bump strength plotted against far-UV steepness for our sample
extinction curves.  The bump strength can be quantified as the area
under the bump using the FM parameters to be $\pi c_3/2 \gamma$
(Fitzpatrick \& Massa 1986).  The steepness of the far-UV extinction
can be characterized using the FM parameter, $c_2$, which represents
the slope of the linear background.  Seven of the nine sightlines
lying in the SD region have the largest values of $c_2$ in our sample.
Six of these sightlines also have bump strengths below the Galactic
average.  The sightlines in our sample outside the SD region have
values of bump strength and $c_2$ grouped around the Galactic average.
As can be seen in Figure 3, the SD region extinction parameters fall
roughly along a line running from the average Galactic values through
the LMC and LMC 2 averages to the those seen in the SMC.  This is a
promising result as it indicates that whatever factors are affecting
the dust properties in the Magellanic clouds may also be affecting the
low density dust in the Milky Way.  As discussed below, the two
SD-region stars which appear in the upper left-hand corner of Figure 3
do not share the extinction properties of the rest of the SD region
sample.

\vspace*{0.05in}
\figurenum{4}
\begin{center}
\plotfiddle{f4.eps}{2.7in}{0}{65}{65}{-200}{-110}
\end{center}
\figcaption{The sightlines marked with circles have weak bumps and steep far-UV 
extinction. The sightlines marked with squares have stronger bumps and less 
extreme far-UV extinction. Filled circles have N(Ca II)/N(Na I) $>$ 0.62 and 
$c_2 >$ 1.0, open circles have N(Ca II)/N(Na I) $<$ 0.63 and $c_2 >$ 1.0, 
filled squares have N(Ca II)/N(Na I) $>$ 0.62 and $c_2 <$ 1.0, open squares 
have N(Ca II)/N(Na I) $<$ 0.63 and $c_2 <$ 1.0,   The wedge represents the 
SD region where forbidden gas velocities were measured by Sembach \& Danks 
(1994).}
\vspace*{0.2in}

\vspace*{0.05in}
\figurenum{5}
\begin{center}
\plotfiddle{f5.eps}{2.7in}{0}{65}{65}{-200}{-110}
\end{center}
\figcaption{Height above or below the Galactic plane plotted against
distance in the plane for the stars in the SD region in Figure 4.  The
symbols are the same as for Figure 4. See text.}
\vspace*{0.2in}

The locations of the stars in our sample are plotted in Figure 4.  We
have separated sightlines according to the steepness of the far-UV
extinction as measured by $c_2$ and also the ratio, N(Ca II)/N(Na I),
which measures the relative fraction of cloud and intercloud medium
along a sightline.  A high value indicates a high fraction of
intercloud medium.  All of the sightlines in Table 1 with both N(Ca
II)/N(Na I) $>$ 0.62 and $c_2 >$ 1.02 lie in the SD region.  Figure 5
shows height above or below the Galactic plane plotted against
distance in the plane for the stars in the SD region.  Figures 4 and 5
clearly show the anomalous extinction is seen only for stars in one
particular direction.  The two stars in the SD region, that do not
share the anomalous extinction of the other sightlines, provide
information on the location of this dust.  HD 151805 has b= 1.59$^o$
so its sightline does not go below the Galactic plane as do the other
stars in the sample.  HD 161653 lies at a distance of 1.8 kpc and is,
by far, the closest star in the l=325$^o$ to 0$^o$ region.  So the
dust responsible for the Magellanic-cloud-like extinction lies further
than about 2 kpc and in a direction defined by 325$^o \leq l \leq 0^o$
and -5$^o \geq b \geq -11^o$.  In fact, the four sightlines with the
steepest extinction, HD 158243, 160993, 163522, and 164340, most
resembling the SMC extinction, lie in an even smaller region bounded
by 337$^o \leq l \leq 352^o$ and -8$^o \geq b \geq -11^o$.  These four
stars also have the lowest reddenings. So the remaining three
less-extreme sightlines are likely a combination of clouds including
more CCM-like dust as well.  Kennedy, Bates, \& Kemp (1998) give a
nice analysis of the absorption components in this direction of the
sky including the sightline to HD 163522 (l = 349.6, b = -9.1, d= 9.4
kpc).  In their picture, nearby clouds at 50 pc, 100 pc to 1 kpc, the
Sagittarius (1.5-2 kpc) and Scutum-Crux (3 kpc) spiral arms all
provide absorption components with negative or small positive
velocities.  They identify peculiar velocity gas corresponding to the
forbidden velocities of +10-50 km s$^{-1}$ found by Sembach \& Danks
(1994).  These forbidden velocities differ significantly from those
expected purely from Galactic rotation.  The origin and distance of
this gas is not well known.  Higher ionization ions are associated
with this gas indicating that it may be associated with a Galactic
fountain or worm (Savage, Massa, \& Sembach 1990; Savage, Sembach, \&
Cardelli 1994).
 
\begin{figure*}
\figurenum{6}
\plottwo{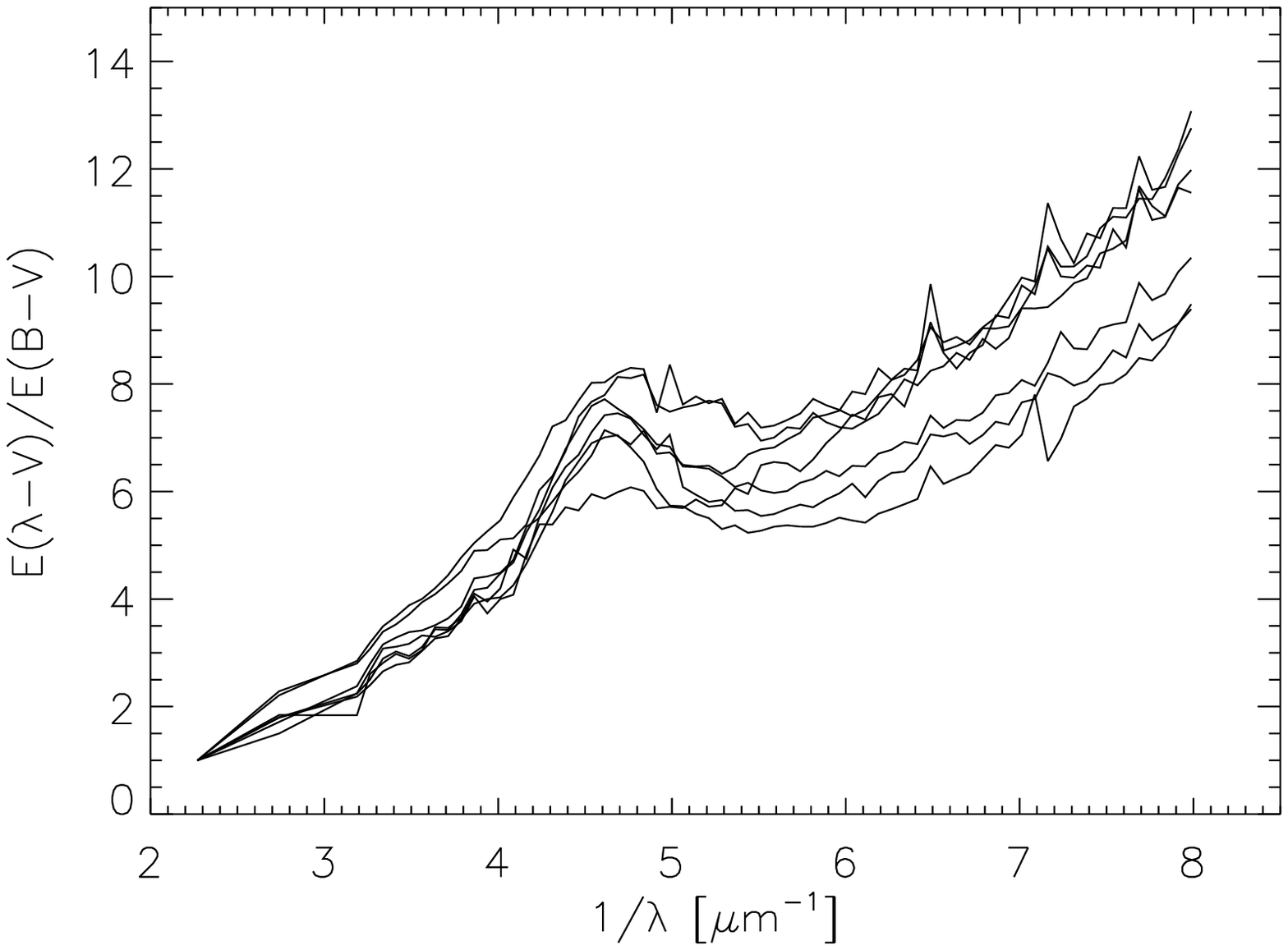}{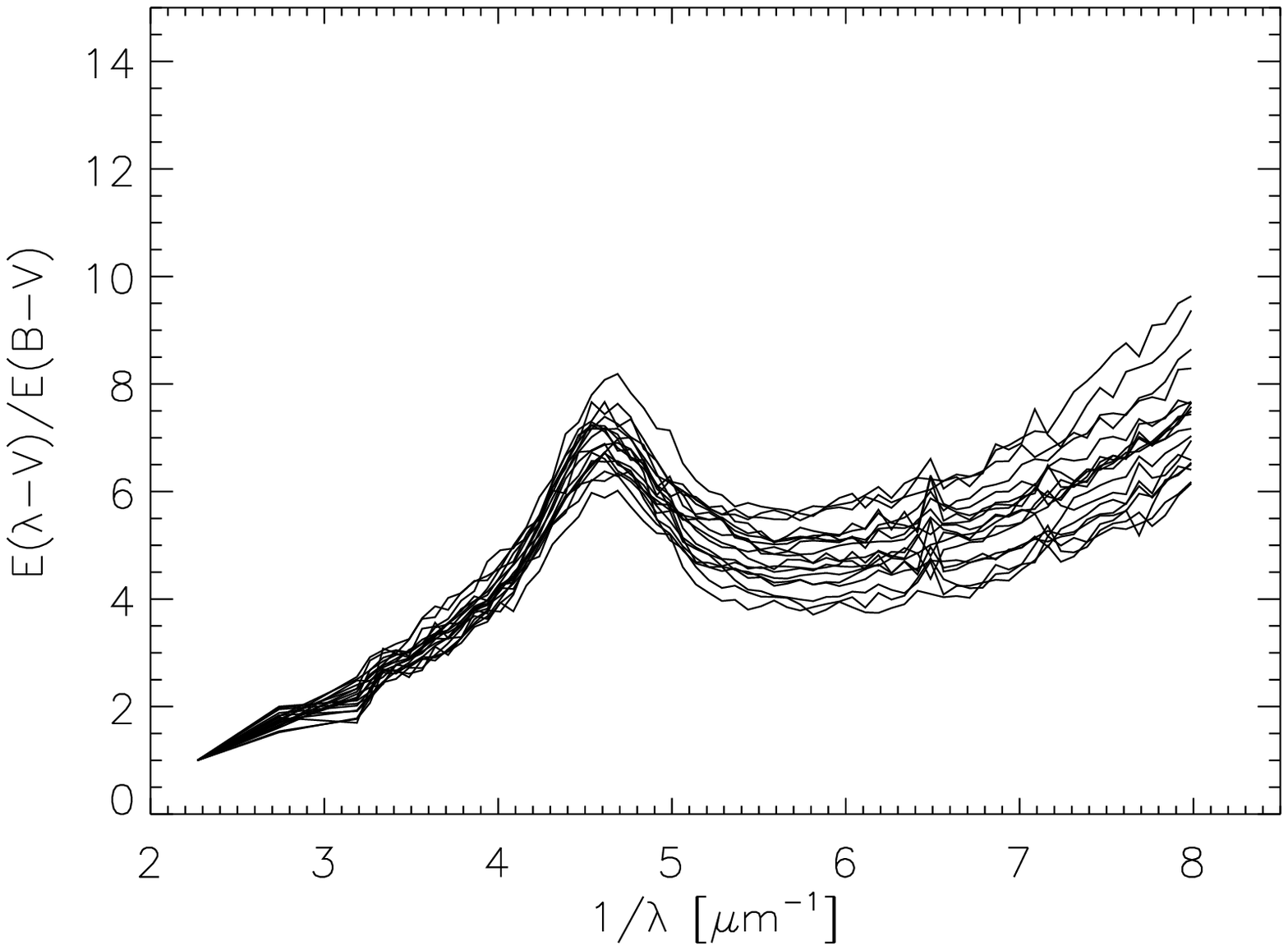}
\caption{The extinction curves for the stars in the SD region (left) and
in the rest of the sample (right) are plotted.}
\end{figure*}

The extinction curves for the sightlines inside and outside of the SD
region are plotted together in Figure 6.  Two sightlines, toward HD
151805 and 161653, in the SD region show normal CCM ($R_V$ = 3.1)
extinction.  As discussed above, the dust along those sightlines is
not located in the same volume of space with the SD type dust so they
are plotted with the non-SD-region sample in Figure 6. The average
SD-region curve (again not including HD 151805 and 161653) is plotted
in Figure 7 along with the average Milky Way and Magellanic cloud
curves for comparison.  The average SD-region curve most resembles the
HD 62542 and 210121 curves, seen in Figure 1, as well as the LMC 2
curve.  The four steepest SD-region curves resemble the SMC Bar curve
in the far-UV but still have stronger bump strengths.  These curves
show little or no slope change from the near to the far-UV while most
other curves seen in Figures 1 and 7 tend to turn up steeply to the
blue of the bump.

The environments of the LMC 2 and HD 62542 dust may be quite similar
to the SD region dust.  All are in diffuse regions subject to shocks
and strong UV radiation fields.  The sightline to HD 210121 contains
one quiescent cloud with E(B-V) =0.40.  However, this cloud is located
in the halo and shows UV extinction quite similar to the that seen in
the SD region. Calcium is heavily depleted indicating that grain
destruction has not been an important mechanism in producing the
unusual extinction (Welty \& Fowler 1992).  The low optical depth of
the Magellanic sightlines implies that the dust is not well shielded
from these environmental pressures.  The typical molecular cloud in
the Magellanic clouds is bigger but more diffuse than in the Galaxy
(Pak et al.  1998).  Then, the small size of the dust grains in a
cloud could be the result of the lack of a very dense environment
necessary for the grains to grow through coagulation.  This has been
suggested as the cause of the anomalous extinction seen along the HD
210121 sightline (Larson et al.  1996; 2000).  This kind of low
density sightline may mimic the conditions in the SMC where extinction
properties have been measured over long sightlines with low values of
E(B-V) = 0.15-0.24 (Gordon \& Clayton 1998).  The value of N(Ca
II)/N(Na I) is not known for the SMC sightlines but the Ca II
abundance in the gas phase is much higher than in the Galaxy for a
given reddening (Cohen 1984).  The gas-to-dust ratio in general in the
SMC is ten times that of the Galaxy (Koornneef 1983).  The gas-to-dust
ratio in our low density Galactic sample is not significantly
different from the average Galactic value (Sembach \& Danks 1994).
This implies that the gas-to-dust ratio is not well correlated with
dust extinction properties.

\vspace*{0.05in}
\figurenum{7}
\begin{center}
\plotone{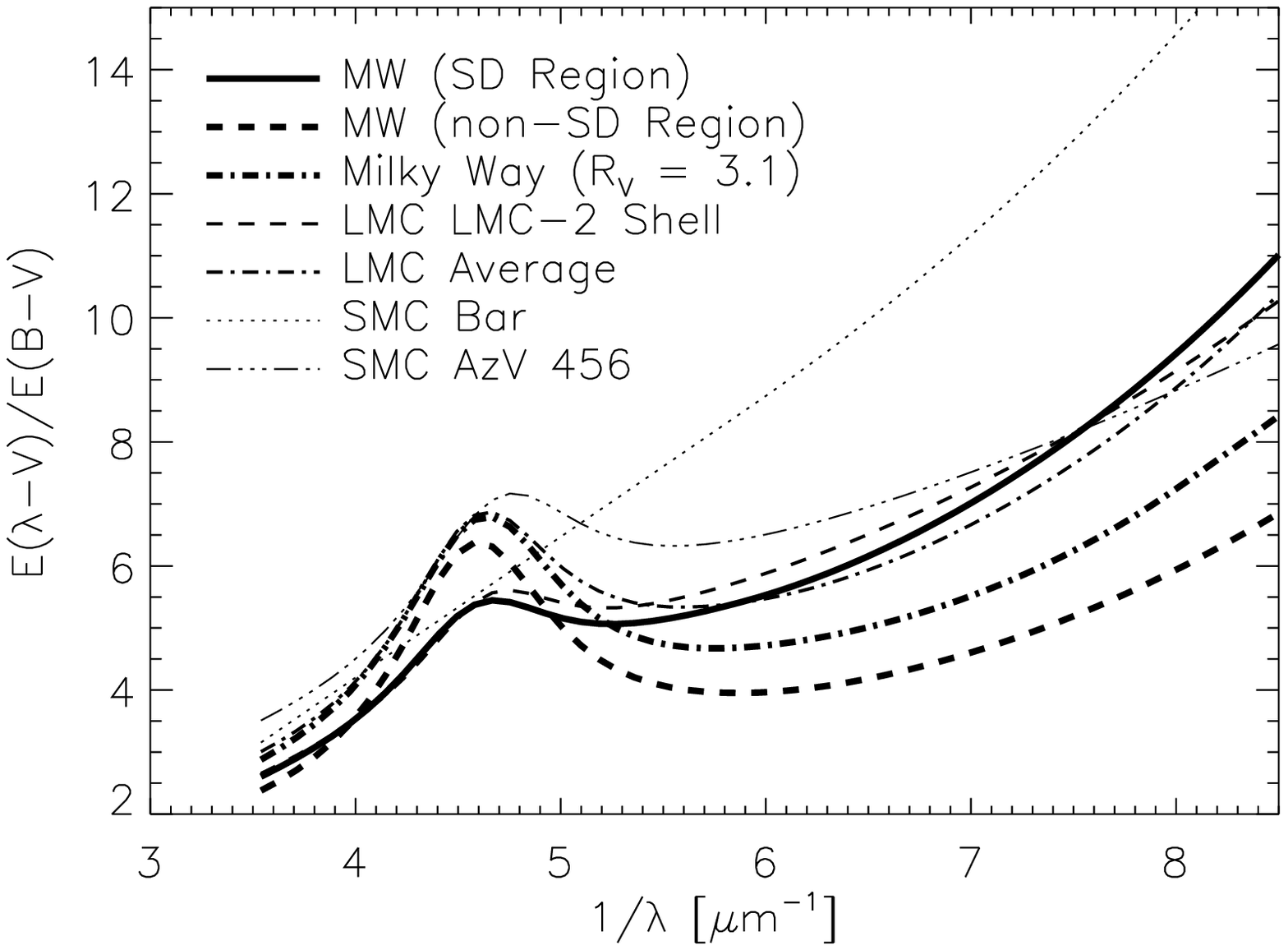}
\end{center}
\figcaption{Comparison of the average SD region and non-SD region curves with
curves for the average MW (R$_V$ = 3.1), LMC Average, LMC 2 Shell, SMC
Bar, and SMC AzV 456.}
\vspace*{0.2in}

The forbidden velocities seen in our sample are associated with warm
intercloud material and the turbulent ISM (Sembach \& Danks 1994) and
may indicate that the dust may have been subject to shocks.  Sembach
\& Savage (1996) investigated the gas and dust abundances in the halo,
finding that they are consistent with progressively more severe
processing of grains from the disk into the halo.  In addition, they
find that while some material is returned to the gas phase, the grain
cores seem resistant to destruction.  Dust models indicate that the
far-UV rise in the extinction curve becomes steeper with increased
frequency of exposure to shocks which produces more small dust grains
(O'Donnell \& Mathis 1997).  However, the frequently shocked dust
models that produce steeper far-UV extinction also result in stronger
bumps.  So, producing an SMC-like extinction curve is not as simple as
placing dust in a diffuse environment and waiting for a supernova
shock.

The next logical step is to attempt to directly connect grain
properties to their respective sightline environments.  Consequently,
we have included the average SD region extinction curve in a
comprehensive study of dust in the Local Group where our goal is to
explicitly examine the correlation of grain size distributions to
sightline characteristics such as depletion patterns, and radiation
environment (Wolff et al. 2000).

\section{Conclusions}

\noindent
$\bullet$ Magellanic-cloud-like extinction has now been found in the Milky Way.\\
$\bullet$ Large values of N(Ca II)/N(Na I) indicating low depletion are associated
with steep far-UV extinction as measured by $c_2$.\\
$\bullet$ Global metallicity seems not to be a direct factor. Local 
environmental conditions seem to be the most important factor in determining 
dust properties.\\
$\bullet$ Similar UV dust extinction properties have now been seen in the 
Milky Way, the Magellanic clouds, starburst galaxies and in high redshift 
star-forming galaxies.\\
$\bullet$ There may be at least two ways to achieve similar extinction
properties. A lack of dust coagulation has been
suggested for HD 210121 to explain the observed extinction
(Larson et al. 1996). The Galactic SD-region 
properties are closely tied to forbidden velocities indicating that processing of 
the grains in the diffuse ISM resulted in their observed properties. \\
$\bullet$ There seems to be a correlation between decreasing bump strength and
far-UV steepness that includes the Galaxy and the Magellanic Clouds. \\
$\bullet$ All the sightlines contained in CCM lie within 1 kpc of the Sun. As 
this study shows, dust properties are not well mapped even in our own Galaxy. 
There are a larger range of UV extinction parameters seen in the Milky Way than 
implied by CCM.\\

Thanks to Ed Fitzpatrick for providing his IUE IDL procedure. This project was 
originally envisioned by the late Jason Cardelli.

\end{document}